\documentclass[12pt]{article}
\usepackage[dvips]{graphicx}
\usepackage{latexsym}
\parskip=\medskipamount

\def \un{\underline}
\newcommand {\cD}{{\cal D}}

\newcommand {\cF}{{\cal F}}
\newcommand {\cG}{{\cal G}}

\newcommand {\cM}{{\cal M}}
\newcommand {\cN}{{\cal N}}

\newcommand {\cP}{{\cal P}}
\newcommand {\cQ}{{\cal Q}}

\newcommand {\cT}{{\cal T}}

\newcommand {\cW}{{\cal W}}
\newcommand {\cX}{{\cal X}}

\newcommand{\bG}{{\bf G}}

\newcommand{\bK}{{\bf K}}

\newcommand{\bW}{{\bf W}}

\def\a{\alpha}
\def \bi{\bibitem}

\def\b{\beta}

\def\d{\delta}
\def\e{\epsilon}
\def\f{\phi}
\def\g{\gamma}
\def\G{\Gamma}

\def\k{\kappa}
\def\l{\lambda}
\def\m{\mu}
\def\n{\nu}
\def\o{\omega}
\def\p{\pi}
\def\q{\theta}
\def\r{\rho}
\def\s{\sigma}
\def\t{\tau}

\def\x{\xi}
\def\z{\zeta}
\def\D{\Delta}
\def\F{\Phi}
\def\J{\Psi}
\def\L{\Lambda}
\def\O{\Omega}

\def\S{\Sigma}
\def\U{\Upsilon}

\newcommand{\ad}{{\dot{\alpha}}}                           
\newcommand{\bd}{{\dot{\beta}}}                            
\newcommand{\ve}{\varepsilon}                            

\newcommand{\hf}{\frac12}

\newcommand{\vf}{\varphi}
\newcommand{\sect}[1]{\setcounter{equation}{0}\section{#1}}

\newcommand{\be}{\begin{equation}}
\newcommand{\ee}{\end{equation}}
\newcommand{\bea}{\begin{eqnarray}}
\newcommand{\eea}{\end{eqnarray}}
\newcommand{\non}{\nonumber}
\newcommand{\1}{\underline{1}}

\newcommand{\iu}{\underline{i}}
\newcommand{\ju}{\underline{j}}
\newcommand{\ku}{\underline{k}}
\newcommand{\lu}{\underline{l}}
\newcommand{\pu}{\underline{p}}
\newcommand{\qu}{\underline{q}}

\begin{document}

\begin{titlepage}
\thispagestyle{empty}

\begin{flushright}
hep-th/0403240 \\
March,  2004 \\
Revised version: July, 2004\\
\end{flushright}
\vspace{5mm}

\begin{center}
{\Large \bf
Relaxed super self-duality 
and N = 4 SYM \\
at two loops}
\end{center}

\begin{center}
{\large S. M. Kuzenko and   I. N. McArthur}\\
\vspace{2mm}

\footnotesize{
{\it School of Physics, The University of Western Australia\\
Crawley, W.A. 6009, Australia}
} \\
{\tt  kuzenko@cyllene.uwa.edu.au},~
{\tt mcarthur@physics.uwa.edu.au} \\
\end{center}
\vspace{5mm}

\begin{abstract}
\baselineskip=14pt
A closed-form expression is obtained for 
a holomorphic sector of the two-loop low-energy effective 
action  for the $\cN=4$ super Yang-Mills theory on 
its Coulomb branch where the gauge group $SU(N)$
is spontaneously broken to $SU(N-1) \times U(1)$ 
and the dynamics is described by a single Abelian 
$\cN=2$ vector multiplet. In the framework of 
the background-field method, this holomorphic sector 
is singled out by computing the effective action 
for a background $\cN=2$ vector multiplet satisfying 
a relaxed super self-duality condition.  
At the two-loop level, the $\cN=4$ SYM effective action is shown 
to  possess no  $F^4$ term (with $F$ the $U(1)$ field strength), 
in accordance with the Dine-Seiberg 
non-renormalization conjecture hep-th/9705057 and 
its generalized form given in hep-th/0310025.
An unexpected outcome of our calculation 
is that no (manifestly supersymmetric invariant 
generating)  $F^6$ quantum correction occurs at 
two loops. This is in conflict with previous results.
\end{abstract}

\vfill
\end{titlepage}

\newpage
\setcounter{page}{1}

\renewcommand{\thefootnote}{\arabic{footnote}}
\setcounter{footnote}{0}
\sect{Introduction}
A year ago, we developed a manifestly covariant 
multi-loop scheme\footnote{The approach of \cite{KM} is 
a natural generalization of  methods developed
in the past in \cite{Schwinger,DeWitt,BV,Avr}, and 
is equally  applicable for computing 
effective actions in non-supersymmetric gauge 
theories.}
 for computing finite  quantum corrections to 
low-energy effective actions, within the background-field method,
in supersymmetric Yang-Mills theories \cite{KM}.
Since then, we have given two important applications 
\cite{KM2,KM3} of the approach suggested. 
In ref. \cite{KM2}, the two-loop 
Euler-Heisenberg effective action for $\cN=2$ 
supersymmetric QED was derived. 
The subject of \cite{KM3}
was a detailed two-loop scrutiny of the Dine-Seiberg 
non-renormalization conjecture
\cite{DS} that the quantum corrections
with four derivatives (or supersymmetric $F^4$ terms) 
are one-loop exact on the Coulomb branch of 
$\cN=2, 4$ superconformal 
field theories in  four space-time dimensions.
To test the Dine-Seiberg proposal, in \cite{KM3}
we compared  the two-loop $F^4$ quantum corrections 
in two different superconformal 
theories with the gauge group $SU(N)$: 
(i) $\cN=4$ SYM; 
(ii) $\cN=2$ SYM with $2N$ hypermultiplets
in the fundamental. According to the Dine-Seiberg conjecture, 
these theories should yield identical two-loop $F^4$ contributions
from all the supergraphs  involving quantum hypermultiplets, 
since the pure $\cN=2$ SYM and ghost sectors are identical
provided the same gauge conditions are chosen.
We explicitly evaluated the relevant two-loop  supergraphs and 
observed that the $F^4$ corrections generated  
have different large $N$ behaviour in the two theories under 
consideration, in obvious conflict with \cite{DS}.

Inspired by the AdS/CFT 
correspondence \cite{Maldacena,GKP,Witten},
it was further conjectured in \cite{KM3} that 
the Dine-Seiberg proposal
should  hold for those $\cN=2$ 
superconformal theories which possess supergravity duals,
including the $\cN=4$ SYM theory
(no supergravity dual exists for 
$\cN=2$ $SU(N)$ SYM with $2N$ hypermultiplets
in the fundamental). In order to test this conjecture 
at two loops in the case of $\cN=4$ SYM,
the hypermultiplet results of \cite{KM3} 
should be complemented with the two-loop
$F^4$ contribution from the pure $\cN=2$ 
SYM and ghost sectors. To compute this contribution
is one of the aims of the present work.

Among the primary motivations for the present work 
was the desire to gather more evidence for 
the conjectured correspondence 
\cite{Maldacena,CT,Tseytlin,BPT}
between the D3-brane action in $AdS_5 \times S^5$
and the low-energy action for $\cN=4$ $SU(N)$ 
SYM on its Coulomb branch,  with the gauge group 
$SU(N)$ spontaneously broken to 
$SU(N-1) \times U(1)$. Let us explain the
final form  \cite{BPT} of
this remarkable conjecture 
in some more detail. 

On the SYM side, the conjecture of \cite{BPT}
deals with the $\cN=4$ SYM effective action 
for an Abelian $\cN=4$  vector multiplet
corresponding to  a particular direction 
in the moduli space of vacua specified by the conditions
(here we consider the bosonic sector only)
\be
\cX_i = X_i \, H_0~, \qquad \cF_{ab} = F_{ab} \, H_0~,
\qquad H_0 = 
{1 \over \sqrt{N(N-1)} }\, 
{\rm diag}(N-1, -1, \ldots, -1)~,
\label{bison0}
\ee
where $i=1, 2,\dots, 6$.
Such a background induces 
spontaneous breaking of 
$SU(N)$ to  $SU(N-1) \times U(1)$, 
and therefore it admits a stringy 
interpretation in terms of a D3-brane separated from 
a stack of $(N-1)$ branes.
In an approximation of slowly varying fields, 
when the scalar fields $X_i$ and the $U(1)$
gauge field strength $F_{ab}$  
are considered to be effectively constant,
the $\cN=4$ SYM effective action should have  
the following general  form 
\be
\G~ = ~ {1 \over g^2_{\rm YM} } 
\int{\rm d}^4 x \,
\sum\limits_{l=0}^{\infty} 
f_l ( g^2_{\rm YM}, N) \, 
{ F^{2l +2} \over |X|^{4l}}~, 
\qquad 
|X|^2 = X_i \,X_i ~.  
\label{bison}
\ee
In the planar (large $N$, fixed $\l =g_{\rm YM}^2N$) 
approximation, the functions $f_l$ should  depend 
on $\l$ only, $f_l ( g^2_{\rm YM}, N) \to f_l( \l)$. 
As conjectured in \cite{BPT} on the basis of the 
AdS/CFT correspondence, 
these functions should possess
the following large $\l$  limit:
\be
  f_l( \l)  ~\stackrel{\l \gg 1}{\longrightarrow}~
a_l \,  \l^l~, 
\label{bison2}
\ee
with some coefficients $a_l$.

On the supergravity  side, the conjecture of \cite{BPT}
deals with the action for a D3-brane probe
in the $AdS_5 \times S^5 $ space 
(oriented parallel to the boundary of $AdS_5$),
which in the case  $X_i = {\rm const}$ becomes
\bea 
S &=& - T_3 \int{\rm d}^4 x \,
{ |X|^4 \over Q} \left\{
\sqrt{ - \det \Big( \eta_{ab} + 
{ Q^{1/2} \over |X|^2} \, F_{ab} \Big) }
-1 \right\} \non \\
&=& {1 \over g^2_s } 
\int{\rm d}^4 x \,
\sum\limits_{l=0}^{\infty} 
c_l\,  ( g_sN )^l \, 
{ F^{2l +2} \over |X|^{4l}}~,
\label{bison3}
\eea
where $T_3 = (2\p g_s)^{-1}$, 
$Q=  g_s N / \p$.
The coefficients $c_l$ can be easily computed  by expanding
the Born-Infeld action in powers of $F$.

The AdS/CFT correspondence is known to require 
$g^2_{\rm YM} =  2\p g_s$. According to the conjecture 
of \cite{BPT},  the coefficients $a_l$ 
in (\ref{bison2}) should be 
directly related to the coefficients $c_l$ in 
 (\ref{bison3}).  The simplest possibility 
to satisfy this condition \cite{BPT} is that the functions
of the coupling, $f_l ( g^2_{\rm YM}, N)$,
in front of some of the terms in (\ref{bison}) 
receive contributions only from the particular orders
in perturbation theory,  and are not renormalized by all 
{\it higher-loop} corrections, see \cite{BPT} 
for a more detailed discussion.

Let us now re-iterate the conjecture
of \cite{BPT} in terms of superfields. 
It is well known that
$\cN=4$ SYM can be viewed as 
$\cN=2$ SYM coupled to a hypermultiplet 
in the adjoint representation of the gauge group.
In $\cN=1$ superfield notation, 
the $\cN=2$ Yang-Mills  supermultiplet 
is described by two chiral 
gauge-covariant superfields
$(\F, \,\cW_\a)$ and their conjugates, 
with $\cW_\a$ the field strength of the $\cN=1$ 
vector multiplet. The hypermultiplet is
described by two chiral scalars
$( \cQ, \,\tilde{\cQ})$ and their conjugates.
On the Coulomb branch, one is  interested in the dynamics 
of light degrees of freedom when
$\cQ= \tilde{\cQ}=0$ and the $\cN=2$ vector 
multiplet takes values in the Cartan subalgebra 
of $SU(N)$. Of special interest 
is the low-energy dynamics 
of a single $U(1)$ vector multiplet
corresponding to  a particular direction 
in the moduli space of vacua specified by the conditions
\be
\F = \f \, H_0~, \qquad \cW_\a = W_\a \, H_0~,
\ee
with the generator $H_0$ given in 
(\ref{bison0}).
Let $\G[W,  \f] $ denote the $\cN=4$ SYM 
low-energy effective action for slowly varying fields.
Its generic structure\footnote{The effective action 
of $\cN=4$ SYM is invariant under quantum-corrected 
superconformal transformations \cite{KMT} which differ
from the ordinary linear superconformal  transformations
that leave  the classical action invariant.
The low-energy action (\ref{structure}) 
corresponds to an approximation when
all terms with derivatives of $\f$ and $\bar \f$
are ignored. For such an approximation, 
the quantum-corrected superconformal 
transformations can be shown to reduce to the classical 
ones, and the latter leave invariant the action 
(\ref{structure}). The superfields $\J^2$ and ${\bar \J}^2$
are scalars with respect to the $\cN=1$ superconformal group
\cite{BKT}.}
is \cite{BKT}  
\be
g^2_{\rm YM}\,
\G[W,  \f] = \hf \int{\rm d}^6 z \,W^2
~+~ 
\int{\rm d}^8 z \,{ {\bar W}^2 W^2 \over  {\bar \f}^2\f^2 } 
\, \O(\J^2, {\bar \J}^2)~,
\label{structure}
\ee
where 
\be
{\bar \J}^2 = {1 \over4} D^2 \Big( { W^2  \over  {\bar \f}^2\f^2 } \Big)~,
\qquad 
\J^2 = {1 \over4} {\bar D}^2 \Big( { {\bar W}^2  \over  {\bar \f}^2\f^2 } 
\Big)~,
\label{psi}
\ee
and $ \O$ is a real analytic function.
The  conjecture of \cite{BPT} now implies that, 
in the planar approximation 
(large $N$, fixed $\l = g_{\rm YM}^2 N$)
accompanied by the large $\l$ limit, 
the $\cN=4$ SYM effective action $\G[W,  \f] $ should   
reduce to the superconformal extension \cite{KT}
of the $\cN=1$ supersymmetric Born-Infeld action \cite{CF} 
\be
g^2_{\rm YM}\,
S_{\rm BI} = \hf
\int{\rm d}^6 z \,W^2
~+~ \hf \k 
\int{\rm d}^8 z \,
\frac{
{\bar W}^2 W^2  ({\bar \f}\f )^{-2} }
{ 1 + \hf A + \sqrt{ 1+ A +{\1 \over 4} B^2 }  }~, 
\label{BI}
\ee
where 
\be 
A =  \k \, ( \J^2 + {\bar \J}^2 ) ~,
\qquad B =  \k  \,( \J^2 - {\bar \J}^2 ) ~, 
\ee
with $\k \propto  \l =g_{\rm YM}^2 N$.

The  conjecture of \cite{BPT} is highly nontrivial. 
It implies that the quantum corrections
to a term of the form
$\J^{2n} \,{\bar \J}^{2m} $,  in the Taylor decomposition
of $\O(\J^2, {\bar \J}^2)$, can occur only 
at loop orders not higher than   $L=1+n+m$, 
and the dominant $L$-loop contribution comes 
with a coefficient proportional to 
$(N g_{\rm YM}^2)^L$; the latter coefficient should 
match the one corresponding to the same structure
in the Born-Infeld action.

The first consequence of the conjecture 
is that the $F^4$ quantum correction
\be 
\int{\rm d}^8 z \,{ {\bar W}^2 W^2 \over  {\bar \f}^2\f^2 } 
\ee
is one-loop exact, and its coefficient fixes 
the Born-Infeld coupling constant $\l^2$ in (\ref{BI}). 
Until recently, the Dine-Seiberg 
non-renormalization theorem \cite{DS} was considered 
to be the firm justification for this. As we now understand, 
there is a subtle flaw in the proof of the theorem given in \cite{DS}.
This proof is claimed  to be applicable to 
all $\cN=2$ superconformal theories, 
and implies the absence of $F^4$ quantum corrections 
beyond one loop in such theories. Unfortunately, 
this does not hold at two loops for some $\cN=2$ superconformal 
theories, as was demonstrated in \cite{KM3} 
by explicit calculations. Nevertheless, it is most likely, 
on symmetry grounds, that the 
$F^4$ correction is indeed one-loop exact in 
$\cN=4$ SYM. As will be shown below, 
there is no two-loop $F^4$ correction 
in $\cN=4$ $SU(N)$ SYM (in contrast with $\cN=2$ 
$SU(N)$ SYM with $2N$ hypermultiplets
in the fundamental).

The second consequence of the above conjecture 
is that the $F^6$ structure
\be
\int{\rm d}^8 z \,{ {\bar W}^2 W^2 \over  {\bar \f}^2\f^2 } 
\, \Big( \J^2 ~+~ {\bar \J}^2 \Big)
\label{n=1-F^6}
\ee
does not receive quantum corrections beyond two 
loops,\footnote{No $F^6$ quantum correction occurs 
in the one-loop effective action for 
$\cN=4$ SYM \cite{FT,BKT}.} 
and the corresponding   two-loop coefficient, 
proportional in the large $N$ limit to  
$(N g_{\rm YM}^2)^2$, 
should  coincide with that 
coming from the Born-Infeld action (\ref{BI}).
The problem of computing the two-loop $F^6$ quantum correction 
was recently addressed in \cite{BPT} on the basis of  
the background field formulation 
in $\cN=2$ harmonic superspace \cite{BBKO}.
The authors of \cite{BPT} reported  complete 
agreement between the $F^6$ terms which occur
in the two-loop $\cN=4$ SYM effective action and 
in the supersymmetric Born-Infeld action (\ref{BI}). 

One of the aims of the present work is to
provide an independent calculation of the
two-loop $F^6$  quantum correction in $\cN=4$ 
SYM using the background field formulation 
in $\cN=1$ superspace \cite{GGRS}
and the multi-loop techniques developed in 
 \cite{KM,KM2,KM3}.  Actually, we compute 
a holomorphic sector of the two-loop 
effective action\footnote{Similarly to \cite{BPT},
in this paper we are interested in the 1PI effective 
action for $\cN=4$ SYM on its Coulomb branch.}, 
of which the $F^6$ term is 
simply a special sub-sector.   
The $F^6$ correction
is associated with the  linear homogeneous 
term of the holomorphic function $\O(\J^2, 0)$.
At the one-loop level, this function is 
known to be trivial \cite{BKT}
\be
\O_{\rm one-loop}(\J^2, 0) = 
\O_{\rm one-loop}(0, 0) ~.
\label{Omega-one-loop}
\ee
What happens at two loops?

The main result of the present paper is 
a closed-form expression  for 
$\O_{\rm two-loop}(\J^2, 0)$ in the case 
of $\cN=4$ $SU(N)$ SYM. This function turns
our to be remarkably simple:
\bea
 \O_{\rm two-loop}(\J^2, 0) 
&=& - {4g_{\rm YM}^4 N(N-1) \over (4 \p)^4 }  
\left( {\J^2 \over  e^2} 
-2 \int_{0}^{\infty} {\rm d} s\, s^2 \Big\{ 
{ \J \over 2 e } \coth {s  \J \over 2e} - {1 \over s} 
\Big\}  \, {\rm  e}^{-s} \right)
~, \non \\
e^2 &=&  {N \over N-1}~.
\label{O-two}
\eea
As can be seen, no $F^6$ term is generated since 
$ \O_{\rm two-loop}(\J^2, 0) = O(\J^4)$.
More precisely, the first term in the expression 
for $\O_{\rm two-loop}(\J^2, 0)$, that  is 
$$
- \,{(2(N-1) g_{\rm YM}^2)^2 \over (4 \p)^4 }  \, \J^2 ~,
$$
matches  exactly, in the large $N$ limit,
 the $F^6$ structure  in the 
Born-Infeld action. However, the similar structure 
in  the second term 
in  $\O_{\rm two-loop}(\J^2, 0)$
differs only in sign from the first, 
and hence the total two-loop $F^6$ correction is zero.

To compute quantum corrections to the holomorphic
function $\O(\J^2, 0)$, it is sufficient
(and extremely advantageous, as far as computational
efforts are concerned) to work with covariant supergraphs
in a vector multiplet background satisfying a relaxed 
super self-duality condition \cite{KM4}.
The latter can be defined by 
\be
W_\a \neq 0~, \quad 
D_\a W_\b = 0~, \qquad 
{\bar D}_{(\ad} {\bar W}_{\bd)} \neq 0
\label{N=1-relaxed}
\ee
in the case of $\cN=1$ supersymmetry, or
\be 
D^i_\a W  \neq 0~, \quad 
D_\a^i  D_\b^j W = 0~, \qquad 
{\bar D}^i{}_{(\ad} {\bar D}_{\bd) i}  {\bar W} \neq 0
\label{N=2-relaxed}
\ee
in  the case of $\cN=2$ supersymmetry.
Here $W$ is the chiral 
superfield strength describing 
an $\cN=2$ Abelian vector multiplet. 
Ordinary (Euclidean) super self-duality \cite{Zumino}
corresponds to  setting 
$W=0$ while keeping 
$\bar W$ non-vanishing.
${}$From the point 
of  view of $\cN=1$ supersymmetry, the $\cN=2$ 
vector multiplet strength 
$W$ consists of two $\cN=1$ superfields:
(i) a chiral scalar $\f$; and (ii) the $\cN=1$ 
vector multiplet strength $W_\a$.
The conditions on $W_\a$ which follow from 
 (\ref{N=2-relaxed}) coincide with 
(\ref{N=1-relaxed}). 

In Minkowski space-time,  
the conditions (\ref{N=1-relaxed})
and (\ref{N=2-relaxed}) are purely formal,  
as they are obviously inconsistent 
with the structure of a single real vector multiplet.
Nevertheless, their use is completely legitimate
if we are  only interested  in computing some
special sector of the effective
action with holomorphic structure. 

This paper is organized as follows. 
Section 2 contains the necessary setup
regarding the $\cN=4$ super Yang-Mills  
theory  and its background field
quantization (for supersymmetric 't Hooft gauge)
in $\cN=1$ superspace. 
The structure of the one-loop effective action 
is briefly discussed in section 3.
In section 4 we work out useful 
functional representations for the two-loop supergraphs.
In section 5 we specify the superfield background 
and collect some important group-theoretical 
results,  following \cite{KM3}.
Section 6 is devoted to the evaluation of the
two-loop supergraphs that come 
from the hypermultiplet sector  of the theory.
The two-loop supergraphs from the pure
$\cN=2$ super Yang-Mills and ghost sectors
are evaluated in section 7.
A discussion of the results obtained  is given in 
section 8. In appendix A  the main properties
of the parallel displacement propagator are given.
In appendix B we discuss
simplifications in the structure of the $U(1)$
superfield heat kernel which occur under the relaxed 
super self-duality condition. 
In appendix C we prove two identities 
given in section 4. Finally, 
appendix D contains details of the group-theoretical 
manipulations leading to the expressions 
(\ref{fish-structure-0}) and (\ref{eight-structure-0}).

\sect{\mbox{$\cN = 4$} SYM setup}

${}$From the point of view of $\cN=2$ supersymmetry,
the classical action of the  $\cN=4$  
super Yang-Mills theory, 
$S = S_{\rm SYM} + S_{\rm hyper}$,
consists of two parts: (i) the pure $\cN=2$ SYM action 
(with $g^2 = 2g_{\rm YM}^2$)
\bea
S_{\rm SYM} &=& \frac{1}{g^2}\,{\rm tr}
\left( \int {\rm d}^8 z \, \F^\dagger \F
+  \int {\rm d}^6 z \, \cW^\a \cW_\a 
\right)~;
\label{n=2pure-sym}
\eea
 (ii) the hypermultiplet action
\bea 
S_{\rm hyper} = \frac{1}{g^2}\,{\rm tr}
\left(
 \int {\rm d}^8 z \, ( \cQ^\dagger  \,\cQ 
+  \tilde{\cQ}^\dagger   \tilde{\cQ} )
 - {\rm i}  \int {\rm d}^6 z \, 
\tilde{\cQ} \,[  \F , \cQ ]
- {\rm i}  \int {\rm d}^6 {\bar z}\,  \tilde{\cQ}^\dagger \,
[\F^\dagger ,  \cQ^\dagger ] \right) ~.
\label{hyper}
\eea 
Here $\F = \F^\m (z) T_\m$, $Q = Q^\m (z) T_\m$ 
and $\tilde{Q} = \tilde{Q}{}^\m (z) T_\m$ are {\it covariantly chiral}
superfields in  the adjoint representation 
of the gauge group, with the latter  chosen  
to be $SU(N)$ throughout this paper.
It is assumed in  (\ref{n=2pure-sym}) 
and (\ref{hyper}) that the trace is taken   
in the fundamental
representation of $SU(N)$, $ {\rm tr} ={\rm tr}_{\rm F} $,
with  the corresponding generators
normalized such that  
${\rm tr} \,(T_\m\, T_\n) = \d_{\m \n}$.
The covariantly chiral superfield strength $\cW_\a$ 
is associated with  the gauge  covariant derivatives
\be
\cD_A = (\cD_a, \cD_\a , {\bar \cD}^\ad ) 
= D_A +{\rm i}\, \G_A~, \qquad 
\G_A = \G^\m_A (z) T_\m~, \qquad 
(T_\m)^\dagger = T_\m~, 
\label{gcd}
\ee
where $D_A$ are the flat covariant 
derivatives\footnote{Our $\cN=1$ notation 
and conventions correspond to \cite{BK}.}, 
and $\G_A$ the superfield connection taking its values 
in the Lie algebra of the gauge group.  
In any representation of the gauge group, 
the   gauge covariant derivatives satisfy the following algebra:
\bea
& \{ \cD_\a , \cD_\b \} 
= \{ {\bar \cD}_\ad , {\bar \cD}_\bd \} =0~, \qquad 
\{ \cD_\a , {\bar \cD}_\bd \} = - 2{\rm i} \, \cD_{\a \bd}~, \non \\
& [ \cD_\a , \cD_{\b \bd}] = 2 {\rm i} \ve_{\a \b}\,{\bar \cW}_\bd ~, 
\qquad 
[{\bar \cD}_\ad , \cD_{\b \bd}] = 2{\rm i} \ve_{\ad \bd}\,\cW_\b ~ , 
\non \\
& [ \cD_{\a \ad}, \cD_{\b \bd} ] 
={\rm i}\, \cF_{\a \ad, \b \bd} 
= - \ve_{\a \b}\, {\bar \cD}_\ad {\bar \cW}_\bd 
-\ve_{\ad \bd} \,\cD_\a \cW_\b~. 
\label{N=1cov-der-al}
\eea
The spinor field strengths $\cW_\a$ and ${\bar \cW}_\ad$ 
obey the Bianchi  identities
\be
{\bar \cD}_\ad \cW_\a =0~, \qquad 
\cD^\a \cW_\a = {\bar \cD}_\ad {\bar \cW}^\ad~.
\ee

To quantize the theory, we will use the $\cN=1$ 
background field formulation \cite{GGRS} and 
split the dynamical variables into background 
and quantum ones, 
\bea
 \F ~ \to ~ \F +\vf ~, \qquad \cQ ~ &\to & ~ \cQ +q ~, 
\qquad 
\tilde{\cQ} ~ \to ~ \tilde{\cQ}+ \tilde{q} ~, \non \\
  \cD_\a ~ \to ~ {\rm e}^{-v} \, \cD_\a \, {\rm e}^v~, 
\quad && \quad 
{\bar \cD}_\ad ~ \to ~ {\bar \cD}_\ad~,
\eea
with lower-case letters used for 
the quantum superfields. 
In this paper, we are not interested in the 
dependence of the effective action on 
the hypermultiplet superfields, 
and therefore we set $\cQ = \tilde{\cQ} =0$
in what follows. 
After the background-quantum splitting, 
the action (\ref{n=2pure-sym}) turns into\footnote{To 
simplify the notation,  we set $g^2 =1$ at the intermediate 
stages of the calculation. The explicit dependence 
on the coupling constant will be restored 
in the final expression for the effective action.}
\bea
S_{\rm SYM} &=& 
{\rm tr} 
\int {\rm d}^8 z \, (\F^\dagger  +\vf^\dagger ) \,
{\rm e}^v \, (\F +\vf) \, {\rm e}^{-v}
+ {\rm tr}
 \int {\rm d}^6 z \, \bW^\a \bW_\a 
\equiv S_{\rm scal} + S_{\rm vect}
~,
\label{bqs-vm}
\eea
where 
\bea
\bW_\a &=& - {1\over 8} {\bar \cD}^2 \Big( 
{\rm e}^{-v}\, \cD_\a \,{\rm e}^{v} \cdot 1 \Big)  = \cW_\a\\
&&  - {1\over 8} {\bar \cD}^2 \Big( 
\cD_\a v- \hf [v, \cD_\a v] 
+{1 \over 6} [v, [v, \cD_\a v]] 
-  {1 \over 24} [v, [v, [v, \cD_\a v]]] 
\Big) + O(v^5)~.  \non 
\eea 
The hypermultiplet  action (\ref{hyper}) takes the form 
\bea 
S_{\rm hyper} &=& {\rm tr}
 \int {\rm d}^8 z \, \Big( q^\dagger \, 
{\rm e}^v \,q \, {\rm e}^{-v}
+  \tilde{q}^\dagger \, 
{\rm e}^{v} \, \tilde{q}  \, {\rm e}^{-v} \Big) 
\non \\  
&-&   {\rm tr} \Big( {\rm i}
\int {\rm d}^6 z \, \tilde{q}\, 
[\F + \vf  ,  q]  +  {\rm i}
\int {\rm d}^6 {\bar z}\,  \tilde{q}^\dagger \,
[\F^\dagger  +\vf^\dagger ,  q^\dagger ] \Big)~.
\label{bqs-hyper}
\eea

It is advantageous to use 
$\cN=1$ supersymmetric 't Hooft gauge 
(a special case of the supersymmetric $R_\x$-gauge
introduced in \cite{OW} and further developed in \cite{BBP})
which is specified by the nonlocal gauge conditions 
\bea
-4 \chi \,&= &{\bar \cD}^2 v +
[\F, ({\Box_+})^{-1} {\bar \cD}^2 \vf^\dagger ] 
= {\bar \cD}^2 v +
[\F, {\bar \cD}^2  ({\Box_-})^{-1}  \vf^\dagger ] 
~, \non \\
-4 \chi^\dagger  &= &\cD^2 v -
[\F^\dagger , ({\Box_-})^{-1} \cD^2 \vf ] 
= \cD^2 v -
[\F^\dagger , \cD^2  ({\Box_+})^{-1}  \vf ] ~.
\eea
Here the covariantly chiral d'Alembertian, $\Box_+$,
is defined by
\bea 
\Box_+ &=& \cD^a \cD_a - \cW^\a \cD_\a -\hf \, (\cD^\a \cW_\a)~, 
\quad
\Box_+ \J = {1 \over 16} \, {\bar \cD}^2 \cD^2 \J ~, \quad 
{\bar \cD}_\ad \J =0~,
\eea
for a covariantly chiral superfield $\J$.
Similarly, the  covariantly antichiral 
d'Alembertian, $\Box_-$, is defined by
\bea 
\Box_- &=& \cD^a \cD_a + {\bar \cW}_\ad {\bar \cD}^\ad 
+\hf \, ({\bar \cD}_\ad  {\bar \cW}^\ad)~, 
\quad
\Box_- {\bar \J} = {1 \over 16} \, \cD^2 {\bar \cD}^2  {\bar \J} ~, 
\quad  \cD_\a {\bar \J} =0~,
\eea
for a covariantly antichiral superfield $\bar \J$.
The gauge conditions chosen lead 
to the following Faddeev-Popov ghost action 
\bea 
S_{\rm gh}
&=& {\rm tr}
 \int {\rm d}^8 z \, (\tilde{c} -\tilde{c}^\dagger ) \left\{
L_{v/2} \, (c+ c^\dagger ) 
+  L_{v/2} \, \coth (  L_{v/2} ) ( c- c^\dagger) 
\right\}  \non \\
&-& {\rm tr}  \int {\rm d}^8 z \, \left\{ 
[\tilde{c}, \F] \, (\Box_-)^{-1}[c^\dagger , \F^\dagger + \vf^\dagger]
+ [ \tilde{c}^\dagger , \F^\dagger] \, (\Box_+)^{-1}
[c, \F + \vf ] \right\} ~,
\eea
with $L_X \, Y =[X,Y] $. 
Here the ghost (Grassmann) superfields $c$ and  $\tilde{c}$  
are  background covariantly chiral.  

A convenient gauge-fixing functional is 
\be
S_{\rm gf} = - {\rm tr}
 \int {\rm d}^8 z \, \chi^\dagger \,\chi~.
\ee
Its introduction is accompanied by the  presence 
of  the Nielsen-Kallosh ghost action 
\be
S_{\rm NK}
= {\rm tr}
 \int {\rm d}^8 z \, b^\dagger \,b ~,
\label{third-ghost}
\ee
where the third (Grassmann)  ghost  superfield $b$ 
is background covariantly chiral.  
The Nielsen-Kallosh ghosts lead to 
a  one-loop contribution only. 

The quantum quadratic part of 
$S_{\rm SYM} + S_{\rm gf}$ is 
\bea 
S^{(2)}_{\rm SYM} + S_{\rm gf} 
 &=&  {\rm tr}
 \int {\rm d}^8 z \,
\Big( \vf^\dagger \,\vf 
- [\F^\dagger , [\F, \vf^\dagger ]] \, 
(\Box_+)^{-1}\, \vf \Big) \non \\
&-&\frac{1}{2}\,{\rm tr} \int {\rm d}^8 z \,
v \, \Big(  \Box_{\rm v}  v -[\F^\dagger, [\F, v]] \Big) 
~+~ \dots 
\label{quad-prel}
\eea 
where the dots stand for the terms with derivatives
of the background (anti)chiral superfields 
 $\F^\dagger$ and $\F$.
The vector d'Alembertian,
$\Box_{\rm v}$,  is defined by
\bea 
{\Box}_{\rm v} 
&=& \cD^a \cD_a - \cW^\a \cD_\a +{\bar \cW}_\ad {\bar \cD}^\ad 
\label{vector-box} \\
&=& -\frac{1}{8} \cD^\a {\bar \cD}^2 \cD_\a 
+{1 \over 16} \{ \cD^2 , {\bar \cD}^2 \} 
-\cW^\a \cD_\a -\hf  (\cD^\a \cW_\a) \non \\
&=& 
 -\frac{1}{8} {\bar \cD}_\ad \cD^2 {\bar \cD}^\ad 
+{1 \over 16} \{ \cD^2 , {\bar \cD}^2 \} 
+{\bar \cW}_\ad {\bar \cD}^\ad +\hf({\bar \cD}_\ad {\bar \cW}^\ad ) ~.
\non 
\eea
The quantum quadratic part of $S_{\rm hyper} $ is 
\bea 
S^{(2)}_{\rm hyper} &=& {\rm tr}
 \int {\rm d}^8 z \, \Big( q^\dagger \, q 
+  \tilde{q}^\dagger \,  \tilde{q}   \Big) 
+  {\rm tr} \Big(
\int {\rm d}^6 z \, \tilde{q}\, 
\cM \,q  +  
\int {\rm d}^6 {\bar z}\,  \tilde{q}^\dagger \,
\cM^\dagger \, q^\dagger  \Big)~, 
\label{2-hyper-ac}
\eea
where the mass operator $\cM$ is defined by 
\be
\cM \, \S = - {\rm i} \, [\F , \S]~, 
\ee
for a superfield $\S$ in the adjoint.

In what follows, 
the background superfields will be  chosen  
to satisfy the conditions:
\be
[\F , \F^\dagger ] = 0~, \qquad \cD_\a \F =0~,
\qquad  \cD^\a \cW_\a = 0~. 
\label{back-con-1}
\ee
Some  additional conditions will be imposed on the background
superfields later on. 
Such an on-shell  background configuration is convenient 
for computing those corrections to the effective action 
which do not contain derivatives of $\F$ and $\F^\dagger$.
The above conditions imply that 
the background superfields belong to the Cartan 
subalgebra of the gauge group. 
Now, the action (\ref{quad-prel}) becomes
\bea 
S^{(2)}_{\rm SYM} + S_{\rm gf} 
 &=& {\rm tr}
 \int {\rm d}^8 z \,
\left( 
\vf^\dagger 
(\Box_+)^{-1} 
(\Box_+ - |\cM|^2 ) \vf 
- \hf 
 v( \Box_{\rm v} - |\cM|^2) v \right)~.
\label{2-sym-ac}
\eea
Similarly, the qudratic part of the Faddeev-Popov
ghost action takes the form
\bea 
S^{(2)}_{\rm gh} 
 &=& {\rm tr}
 \int {\rm d}^8 z \,
\left( 
c^\dagger 
(\Box_+)^{-1} 
(\Box_+ - |\cM|^2 ) \tilde{c}
- \tilde{c}^\dagger 
(\Box_+)^{-1}
(\Box_+ - |\cM|^2 ) c \right)~.
\label{2-gh-ac}
\eea

The cubic  and quartic parts of $S_{\rm scal}$ are:
\bea 
S^{(3)}_{\rm scal}  &=& 
 {\rm tr}  \int {\rm d}^8 z \, \left(
\vf^\dagger [v,\vf ] 
+ {1 \over 6} \F^\dagger [v,[v[,v, \F]]] 
+\hf \Big( \F^\dagger [v,[v, \vf]] +{\rm h.c.} \Big)
\right) ~,  \label{scal-3}\\
S^{(4)}_{\rm scal}  &=& 
\hf {\rm tr}  \int {\rm d}^8 z \, \left(
\vf^\dagger [v,[v, \vf ]]  
+ {1 \over 3} 
\Big(  \F^\dagger [v,[v,[v, \vf]]] +{\rm h.c.} \Big)
\right. \non \\
&& \left.
\qquad \qquad \qquad \qquad
+ {1 \over 12} [v,[v, \F^\dagger]]\, [v,[v,\F]] \right)~.
\label{scal-4}
\eea

The cubic  and quartic parts of $S_{\rm vect}$ are:  
\bea 
S^{(3)}_{\rm vect}  
 &=& \phantom{-}
 \hf {\rm tr}
 \int {\rm d}^8 z \, 
[v, \cD^\a v] \, 
\left( {1 \over 8} \, 
{\bar \cD}^2 \cD_\a v 
 + {1 \over 3} 
[\cW_\a, v ]  \right) \equiv 
S^{(3)}_{\rm vect,I}  + S^{(3)}_{\rm vect,II} ~, 
\label{vect-3} \\
S^{(4)}_{\rm vect}  
 &=&  -{1 \over 8}{\rm tr}
 \int {\rm d}^8 z \,  
[v, \cD^\a v] \left( {1 \over 8}
{\bar \cD}^2 [v, \cD_\a v]
- {1 \over 6}  [v,  {\bar \cD}^2 \cD_\a v] 
+ {1 \over 3} [v,[v, \cW_\a ]] \right)~.
\label{vect-4}
\eea
It is an instructive exercise to show, 
using the algebra of covariant derivatives  
(\ref{N=1cov-der-al}), 
that the functionals $S^{(3)}_{\rm vect} $ and
$S^{(4)}_{\rm vect} $ are real modulo total derivatives.
However, it turns out advantageous for loop calculations 
\cite{GZ} to keep these interaction terms in the complex form
given.

The cubic  and quartic parts of $S_{\rm hyper}$ are:  
\bea 
S^{(3)}_{\rm hyper} &=& {\rm tr} \left\{
 \int {\rm d}^8 z \, \Big( q^\dagger \, [v ,q ]
+  \tilde{q}^\dagger \, 
[v,  \tilde{q} ] \Big)
-  {\rm i} \int {\rm d}^6 z \, \tilde{q}\, 
[ \vf  ,  q]  - {\rm i}
\int {\rm d}^6 {\bar z}\,  \tilde{q}^\dagger \,
[\vf^\dagger ,  q^\dagger ] \right\}~ ,
\label{hyper-3}\\
S^{(4)}_{\rm hyper} &=& \hf {\rm tr} 
 \int {\rm d}^8 z \, \Big( q^\dagger \, [v,[v ,q ] ]
+  \tilde{q}^\dagger \, 
[v,[v,  \tilde{q} ] ]\Big)~.
\label{hyper-4}
\eea

The cubic  and quartic parts of $S_{\rm gh}$ are:  
\bea 
S^{(3)}_{\rm gh}
&=& \hf {\rm tr}
 \int {\rm d}^8 z \, (\tilde{c} -\tilde{c}^\dagger ) \,
[v, (c+ c^\dagger ) ] 
  \non \\
&& -{\rm tr}  \int {\rm d}^8 z \, \left\{ 
[\tilde{c}, \F] \, (\Box_-)^{-1}
[c^\dagger ,  \vf^\dagger]
+ [ \tilde{c}^\dagger , \F^\dagger] \, (\Box_+)^{-1}
[c,  \vf ] \right\} ~,  
\label{gh-3} \\ 
S^{(4)}_{\rm gh}
&=& {1 \over 12}  {\rm tr}
 \int {\rm d}^8 z \, (\tilde{c} -\tilde{c}^\dagger ) \,
[v,[v, (c -c^\dagger )]]~.
\label{gh-4}
\eea

${}$For the background chosen, 
the Feynman propagators 
associated with the quadratic actions
(\ref{2-hyper-ac}),  (\ref{2-sym-ac}) and (\ref{2-gh-ac})
can be expressed via a single Green's function.
Such a Green's function,
$G(z,z')$,  originates 
in the following auxiliary  model
\be
S = \hf
 \int {\rm d}^8 z \,
\S^{\rm T}
( \Box_{\rm v} - |\cM|^2) \S~,
\label{S-action}
\ee
which describes the dynamics of an unconstrained 
real superfield $\S =( \S^\m )$ transforming 
in the adjoint representation of the gauge group. 
The relevant Feynman propagator reads
\be
G(z,z') = 
{\rm i}\,\langle 0| \cT \, \Big(\S(z)\,  \S^{\rm T} (z') \Big) |0\rangle
\equiv  {\rm i}\,\langle \S(z) \, \S^{\rm T} (z')\rangle
\ee
and satisfies the equation
\be
\Big(\Box_{\rm v} - |\cM|^2  \Big) 
\, G(z,z') 
= - {\bf 1}\,\d^8 (z-z')~.
\label{green}
\ee
In listing the Feynman propagators
associated with the  actions
(\ref{2-hyper-ac}),  (\ref{2-sym-ac}) and (\ref{2-gh-ac}), 
 all the dynamical adjoint superfields will be treated 
as  column-vectors (e.g.,  $\vf = (\vf^\m)$
and ${\bar \vf} = ({\bar \vf}^\m)$, 
with $\vf^\dagger$ being a row-vector), 
and not as Lie-algebra-valued objects,
say $v=v^\m T_\m$,  
as they  have been understood so far.
The Feynman propagators 
for the action (\ref{2-sym-ac}) are:
\bea
 {\rm i}  \, \langle  v (z)\, v^{\rm T} (z') \rangle &=& 
-   G(z,z') ~, \non \\
{\rm i}  \, \langle  \vf (z)\,  \vf^\dagger (z') \rangle &=& 
{1 \over 16} {\bar \cD}^2 \cD'^2\, 
G(z,z') ~, \qquad
 \langle  \vf (z)\, \vf^{\rm T} (z') \rangle =
\langle  {\bar \vf} (z)\, \vf^\dagger (z') \rangle
=0~. 
\eea
The Feynman propagators 
for the action (\ref{2-hyper-ac}) are:
\bea
 {\rm i}  \, \langle   q  (z)\,  q ^\dagger (z') \rangle &=& 
{\rm i}  \, \langle \tilde{q}  (z)\,  \tilde{q} ^\dagger (z') \rangle
= {1 \over 16} {\bar \cD}^2 \cD'^2\, 
G(z,z') ~, \non \\
{\rm i}  \, \langle  q  (z)\,  \tilde{q} ^{\rm T} (z') \rangle &=& 
\cM^\dagger \,  
\Big(-{1 \over 4} {\bar \cD}^2 \Big) G(z,z') 
= \cM^\dagger \,  
\Big(-{1 \over 4} {\bar \cD}'^2 \Big) G(z,z') ~, \\
{\rm i}  \, \langle  
\overline{\tilde{q}}  (z) \, q ^\dagger (z') \rangle &=&  
\cM \,  \Big(-{1 \over 4} \cD^2 \Big) G(z,z') 
= \cM \,  \Big(-{1 \over 4} \cD'^2 \Big) G(z,z') 
 ~.\non  
\eea
The Feynman propagators 
for the action (\ref{2-gh-ac}) are:
\bea
 {\rm i}  \, \langle   \tilde{c}  (z)\,  c^\dagger (z') \rangle =
- {\rm i}  \, \langle c  (z)\,  \tilde{c} ^\dagger (z') \rangle 
= {1 \over 16} {\bar \cD}^2 \cD'^2\, 
G(z,z') ~.
\eea

\sect{One-loop effective action}

The one-loop contribution to the effective action 
is determined by the quantum quadratic actions
(\ref{third-ghost}), (\ref{2-hyper-ac}),  (\ref{2-sym-ac})
and (\ref{2-gh-ac}). It can be shown that the contributions 
coming from the quantum chiral superfields  
(i.e. $b$, $c$, $\tilde{c}$, $q$,  $\tilde{q} $ and $\vf$) 
and their conjugates,  
cancel each other\footnote{The relation 
(\ref{one-loop-ea}) was first  derived in \cite{BBK}
using the background field formulation in 
$\cN=2$ harmonic superspace.} 
(see the second reference in \cite{BBP}).
As a result, the one-loop effective action is 
solely generated  by the contribution 
coming from the quantum gauge superfield 
in (\ref{2-sym-ac}):
\be
\G_{\rm one-loop} = -{ {\rm i} \over 2} \,{\rm Tr} \, 
\ln  \, G = - { {\rm i} \over 2} \int\limits_{0}^{\infty} 
{ {\rm d}s \over s} \, 
{\rm Tr} \, \Big\{ K(s) \, 
{\rm e}^{ -{\rm i} (|\cM|^2 -{\rm i}\ve ) s } \Big\}~.
\label{one-loop-ea}
\ee
Here $K(s) \,
\exp (
-{\rm i}\, |\cM|^2  \,s ) $ 
is the heat kernel associated with 
the Green's function $G$ satisfying 
 the equation (\ref{green}) and 
the Feynman boundary conditions, 
\be
G(z,z') = 
{\rm i} \int\limits_0^\infty {\rm d}s \, K(z,z'|s) \, 
{\rm e}^{ -{\rm i} (|\cM|^2 -{\rm i}\ve ) s }~, 
\qquad   \ve \to +0~.
\label{proper-time-repr}
\ee
The functional trace in (\ref{one-loop-ea}) is defined 
as follows 
\be 
{\rm Tr} \, \Big\{ K(s) \, 
{\rm e}^{ -{\rm i} (|\cM|^2 -{\rm i}\ve ) s } \Big\}
= {\rm tr}_{\rm Ad} \int {\rm d}^8 z \, 
K(z,z|s)\, {\rm e}^{ -{\rm i} (|\cM|^2 -{\rm i}\ve ) s } ~,
\ee
where ${\rm tr}_{\rm Ad}$ denotes the operation of trace 
in the adjoint representation of the gauge group.
The background superfields are chosen to satisfy 
the conditions (\ref{back-con-1}).

${}$For  the case of a covariantly 
constant background vector multiplet,
\be
 \cD^\a \cW_\a = 0~, \qquad
\cD_a \cW_\b =0~, 
\label{back-con-2}
\ee
the exact expression for the heat kernel 
is known\footnote{This
heat kernel was first found in 
Fock-Schwinger 
gauge in \cite{O}.}
\cite{KM,KM2}:
\bea
K(z,z'|s) &=& -\frac{\rm i}{(4 \pi s)^2} \, 
\sqrt{
{\bf  det}
\left( \frac{s \,\cF}{\sinh  (s \cF )}\right) } 
\; {\rm U}(s) \,
\z^2  \bar{\z}^2 \,
{\rm e}^{ \frac{{\rm i}}{4} 
\r \, \cF \coth ( s \cF) \, \r } \, I(z,z') 
\label{real-kernel} \\
&=& -\frac{\rm i}{(4 \pi s)^2} \, 
\sqrt{
{\bf det}
\left( \frac{s \,\cF}{\sinh  (s \cF )}\right) } 
\;  \z^2(s)   \bar{\z}^2 (s) \,
{\rm e}^{ \frac{{\rm i}}{4} 
\r (s) \, \cF \coth ( s \cF) \, \r (s) } 
\, {\rm U}(s) \,I(z,z') ~,  \non  
\eea
where the determinant is computed
with respect to the Lorentz indices,
\be 
{\rm U}(s) = \exp \Big\{- {\rm i} s (\cW^{\a} \cD_{\a} 
+ \bar{\cW}^{\ad} {\bar \cD}_{\ad})\Big\}~,
\ee
and $I(z,z') $ is the so-called parallel displacement propagator, 
see  Appendix A for its definition and basic properties. 
The supersymmetric two-point function 
$\z^A(z,z') =- \z^A(z',z)=(\r^a , \z^\a, {\bar \z}_\ad)$
is defined as follows: 
\be
\r^a = (x-x')^a - {\rm i} (\q-\q') \s^a {\bar \q}' 
+ {\rm i} \q' \s^a ( {\bar \q} - {\bar \q}') ~, \quad
\z^\a = (\q - \q')^\a ~, \quad
{\bar \z}_\ad =({\bar \q} -{\bar \q}' )_\ad ~. 
\label{two-point}
\ee
With the notation\footnote{The symbol 
$\bf tr$ denotes the trace with respect to spinor indices.}
\be
\cN_\a{}^\b = \cD_\a \cW^\b~, \qquad 
{\bar \cN}_\ad{}^\bd = {\bar \cD}_\ad {\bar \cW}^\bd~, 
\qquad {\bf  tr} \, \cN = {\bf tr} \, {\bar \cN} =0~,
\ee
for special proper-time dependent variables of 
the form $\J(s) \equiv {\rm U}(s) \, \J \, {\rm U}(-s)$
one gets
\bea
 \cW^{\a}(s) 
= (\cW\, {\rm e}^{- {\rm i} s \cN} )^{\a}~, 
\quad && \quad
{\bar \cW}^{\ad}(s) 
= ({\bar \cW}\, {\rm e}^{- {\rm i} s {\bar \cN}} )^{\ad}~, 
\nonumber \\
 \z^{\a}(s) 
= \z^{\a} + \Big( \cW \,
\frac{  {\rm e}^{ -{\rm i} s \cN} -1} {\cN} \Big)^{\a}~,
\quad && \quad 
{\bar \z}^{\ad}(s) 
= {\bar \z}^{\ad} - \Big( {\bar \cW} \,
\frac{  {\rm e}^{-{\rm i}s {\bar \cN}} -1} {\bar \cN}\Big)^{\ad}~, 
\label{zeta(s)}  \\
\r_{\a \ad}(s) 
= \r_{\a \ad} &-&2 \int_0^{s} {\rm d}t \, 
\Big( \cW_{\a}(t) \bar{\z}_{\ad}(t) 
+ \z_{\a}(t)\bar{\cW}_{\ad}(t) \Big)~.
\non
\eea
Here $\r^a(s) $,  $ \z^{\a}(s) $ and  ${\bar \z}_{\ad}(s) $ 
are the building blocks which appear in  the second line of
(\ref{real-kernel}).
The explicit expression for ${\rm U}(s) \, I(z,z')$
is given in \cite{KM}. 

The parallel displacement propagator is the only 
building block for the supersymmetric heat kernel
which involves the naked gauge connection.
In covariant supergraphs, however, 
the parallel displacement propagators 
that come from all possible internal lines 
`annihilate' each other through the mechanism 
sketched in \cite{KM}. At one loop, 
the parallel displacement propagator 
disappears because of  the identity  (\ref{super-PDO3}).
At two and higher loops, it is the identity 
(\ref{collapse}) and its generalizations
which are responsible for annihilation 
of the parallel displacement propagators.

Since the background fields take their values in
the Cartan subalgebra, then
\be 
\cW_\a \, \cW_\b \, \cW_\g = 0~, \qquad 
{\bar \cW}_\ad \, {\bar \cW}_\bd \,
{\bar \cW}_{\dot{\g}} = 0~,
\label{three-then-zero}
\ee
and therefore the heat kernel at coincident 
points is
\bea
K(z,z|s) &=& -\frac{\rm i}{(4 \pi s)^2} \, 
\sqrt{
{\bf  det}
\left( \frac{s \,\cF}{\sinh  (s \cF )}\right) } \;
{\bf tr} \left( {\sin^2(s \cN/2) \over (\cN/2)^2} \right)\,
{\bf tr} 
\left( {\sin^2(s {\bar \cN}/2) \over ({\bar \cN}/2)^2} \right) \non \\
 & & \qquad \times {1\over 4}\,  \cW^2 \, {\bar \cW}^2~.
\eea
Introducing the (anti) self-dual components of $\cF$, 
\be 
\cF_\pm = \hf (\cF \mp {\rm i}\, \tilde{\cF})~,
\qquad \widetilde{\cF_\pm } = \pm \,{\rm i} \, \cF_\pm ~,
\ee
with $\tilde{\cF}$ the Hodge-dual of $\cF$, 
the above result can be rewritten as follows
\bea
K(z,z|s) &=& -\frac{\rm i}{(4 \pi s)^2} \, 
\sqrt{
{\bf  det}
\left( \frac{s \,\cF}{\sinh  (s \cF )} \,
\frac{ \sinh (s\cF_+)}{\cF_+} \,
\frac{ \sinh (s\cF_-)}{\cF_-} 
\right) } \;
\cW^2 \, {\bar \cW}^2~.
\eea

Imposing the condition of relaxed super self-duality
discussed  in the Introduction, 
\be
\cF_- =0~,  \qquad \cF_+ \neq 0~,
\ee
the expression for $K(z,z|s)$ simplifies drastically, 
\bea
K(z,z|s) &=& -\frac{{\rm i}\,s^2}{(4 \pi )^2} \, 
\cW^2 \, {\bar \cW}^2~.
\eea
The result (\ref{Omega-one-loop}) follows from this.
${}$If the $\cN=2$ vector multiplet is aligned along 
a generic direction in the moduli space of vacua
(with the Cartan-Weyl basis for $SU(N)$ specified
 in eqs.  (\ref{C-W}) and (\ref{C-W-2})), 
\be
\F = \f^I \,H_I ~, \qquad 
\cW_\a = W^I_\a \, H_I~, 
\ee
then one can show that the one-loop effective 
action coincides with the result given in 
\cite{BBK} (see also \cite{G-R}). For a special  
field configuration, eq. 
(\ref{actual-background}),  that corresponds 
to the spontaneous breakdown of $SU(N)$ to 
$SU(N-1) \times U(1)$,  one obtains
\be
\G_{\rm one-loop} =
\frac{N-1}{(4 \pi )^2} 
\int{\rm d}^8 z \,{ {\bar W}^2 W^2 \over  {\bar \f}^2\f^2 } ~,
\ee
and therefore
\be
\O_{\rm one-loop}(\J^2, 0)=
\frac{g_{\rm YM}^2  (N-1)}{(4 \pi )^2} ~.
\ee

\sect{Functional representation for two-loop supergraphs}
We now turn to obtaining a useful functional representation 
for the two-loop supergraphs. Most of the consideration of 
this section is valid for an arbitrary gauge group. 

Even  a quick glance at the structure of the cubic and 
quartic interactions is sufficient to recognize that
the supergraphs generated by 
the pure $\cN=1$ SYM vertices (\ref{vect-3}) and 
(\ref{vect-4}) are the most laborious ones.
Their evaluation  turns out to 
simplify significantly if the background 
superfields are subject to an additional 
restriction of the form
\be 
\cD_\a \cW_\b = 0~, \qquad \quad 
{\bar \cD}_{(\ad} {\bar \cW}_{\bd )} \neq 0~.
\label{hol-bak}
\ee 
As discussed in the Introduction,
these conditions are rather formal, 
since they are incompatible 
with a real vector supermultiplet in Minkowski space. 
However,  
their use is completely legitimate if 
we are only interested in computing a special 
holomorphic sector of the effective action.

To explain why the first condition in (\ref{hol-bak})
is useful, let us point out that without this condition 
the heat kernel satisfies the identity 
\be 
\cD'_\a \, K(z,z'|s) 
= - ( {\rm e}^{ {\rm i} s \cN } )_\a{}^\b \, \cD_\b 
\, K(z,z'|s)~, \qquad 
\cN_\a{}^\b = \cD_\a \cW^\b~.
\ee
Therefore
\be 
\cD_\a \cW_\b = 0 \quad \longrightarrow
\quad 
\cD'_\a \, K(z,z'|s) 
= -  \cD_\a \, K(z,z'|s)~,
\ee
and similarly for the corresponding Green's function, 
\be
\cD_\a \cW_\b = 0 \quad \longrightarrow
\quad 
\cD'_\a \, G(z,z') 
= -  \cD_\a \, G(z,z')~.
\label{A1}
\ee
The latter identity proves to be  invaluable 
when evaluating  the supergraphs generated by 
the pure $\cN=1$ SYM vertices (\ref{vect-3}) and 
(\ref{vect-4}). 

\begin{figure}[!htb]
\begin{center}
\includegraphics{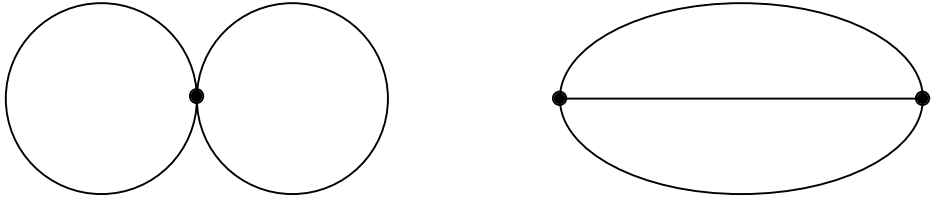}
\caption{Two-loop supergraphs: `eight' diagram and `fish' diagram.}
\end{center}
\end{figure}

\subsection{Hypermultiplet  `fish' supergraphs}
The two-loop supergraphs generated by the 
hypermultiplet vertices (\ref{hyper-3}) and 
(\ref{hyper-4}) have been analyzed in detail in 
\cite{KM3}, and therefore we simply 
reproduce the results.

The cubic interaction  (\ref{hyper-3})
generates the following 
fish-type contributions 
to the effective action 
\bea
{ {\rm i} \over 2}  \, \Big\langle S^{(3)}_{\rm hyper} \, 
 S^{(3)}_{\rm hyper} \Big\rangle_{\rm 1PI}  
= \G_{\rm I} +  \G_{\rm II}~,
\eea 
where 
\bea
 \G_{\rm I} &=& {1 \over 2^{9}} 
 \int {\rm d}^8 z   \int {\rm d}^8 z' \, 
G^{\m \n}(z,z') 
{\rm tr}_{\rm Ad} \Big(
 T_\m \, [ {\bar \cD}^2 ,  \cD^2 ]  G(z,z') 
\, T_\n \, [ {\bar \cD}'^2 , \cD'^2 ]  G(z',z) \Big) ~, \non \\
 \G_{\rm II} &=& -
{ 1 \over 2^4} \int {\rm d}^8 z   \int {\rm d}^8 z' \, 
G^{\m \n}(z,z') {\rm tr}_{\rm Ad} \Big(
 T_\m \, \F^\dagger \, {\bar \cD}^2   G(z,z') 
\, T_\n \, \F \,   \cD'^2   G(z',z) \Big)~.
\label{hyper-fish}
\eea
The quantum corrections
$ \G_{\rm I} $ and $ \G_{\rm II} $ 
were denoted in \cite{KM3}
as $\G_{\rm I+II}$ and $\G_{\rm III}$, 
respectively.

\subsection{Hypermultiplet  `eight' supergraphs}
The quartic interaction  (\ref{hyper-4})
generates an `eight' supergraph, 
\be 
\Big\langle S^{(4)}_{\rm hyper}   
\Big\rangle_{\rm 1PI}  
\equiv  \G_{\rm III} ~,
\ee 
which has the following structure \cite{KM3}
\bea
 \G_{\rm III}  
= {1 \over 2^4 }   \int {\rm d}^8 z  
\lim_{z' \to  z} \,
G^{\m \n}(z,z') \,{\rm tr}_{\rm Ad}
 \Big( T_\m \, {\bar \cD}^2\cD^2 G(z,z') \, T_\n \Big) ~.
\label{hyper-eight}
\eea
This quantum correction was 
denoted in \cite{KM3} as $\G_{\rm IV}$.

\subsection{Vector `fish' supergraphs}
The cubic interaction  (\ref{vect-3})
generates the following 
fish-type contributions 
to the effective action 
\bea
{ {\rm i} \over 2}  \, \Big\langle S^{(3)}_{\rm vect,I} \, 
 S^{(3)}_{\rm vect,I} \Big\rangle_{\rm 1PI}  
~+~  {\rm i}  \, \Big\langle S^{(3)}_{\rm vect,I} \, 
 S^{(3)}_{\rm vect,II} \Big\rangle_{\rm 1PI}  
\equiv \D \G_{1} + \D \G_{2}~,
\eea
where
\bea
\D \G_{1} &= &
 {1 \over 2^{10}} 
 \int {\rm d}^8 z   \int {\rm d}^8 z' \, 
G^{\m \n}(z,z') \,{\rm tr}_{\rm Ad} \,T_\m    
\Big( 2   \cD^2 G(z,z') \, T_\n \,
{\bar \cD}'^2 \cD'^2 {\bar \cD}'^2 G(z',z) \non \\
&+&   [ {\bar \cD}^2 , \cD^2 ] G(z,z') \, 
T_\n \, \cD'^\a {\bar \cD}'^2 \cD'_\a G(z',z) 
-   \, \cD^2 {\bar \cD}^2 G(z,z') \,T_\n \,
\{   {\bar \cD}'^2 , \cD'^2\} G(z',z) \non \\
&-&   \cD^\a G(z,z') T_\n \,
{\bar \cD}'^2 \cD'^2 {\bar \cD}'^2 \cD'_\a G(z',z)
 \Big) ~, 
\label{vector-fish-1.a}\\
\D \G_{2} &=& 
{1 \over 3 \cdot 2^5} 
 \int {\rm d}^8 z   \int {\rm d}^8 z' \, 
G^{\m \n}(z,z') \,{\rm tr}_{\rm Ad}  \,T_\m
\Big(   \cD^2 G(z,z') \, \{ T_\n , \cW'^\a \} \, 
\cD'_\a {\bar \cD}'^2 G(z',z)  \non \\
&-&  \cD^\a G(z,z')  \, \cW'_\a \, T_\n \,
\cD'^2 {\bar \cD}'^2 G(z',z) 
- \hf  \cD^\a G(z,z') \, T_\n  \, \cW'_\a \, 
\cD'^2 {\bar \cD}'^2 G(z',z) \Big) ~.  
\label{vector-fish-2.a}
\eea 
One can readily show that 
\be
\Big\langle S^{(3)}_{\rm vect,II} \, 
 S^{(3)}_{\rm vect,II} \Big\rangle_{\rm 1PI}  =0
\ee
for the background  configuration, 
specified in the next section, 
eq. (\ref{actual-background}).

The expression for $\D \G_{2}$ can be brought to 
a simpler form, 
\bea
\D \G_{2} &=& 
{1 \over 2^6} 
 \int {\rm d}^8 z   \int {\rm d}^8 z' \, 
G^{\m \n}(z,z') \,{\rm tr}_{\rm Ad} 
\Big(  T_\m\, \cD^2 G(z,z') \,  T_\n \, \cW'^\a  \, 
\cD'_\a {\bar \cD}'^2 G(z',z)  \non \\
&-& T_\m\,  \cD^\a G(z,z')  \, \cW'_\a \, T_\n \,
\cD'^2 {\bar \cD}'^2 G(z',z) \Big) ~,
\label{vector-fish-2.b}
\eea
if one notices the following identities
\bea
0 &=&  \int {\rm d}^8 z   \int {\rm d}^8 z' \, 
G^{\m \n}(z,z') \,{\rm tr}_{\rm Ad}  \Big( 
T_\m  \, \cD^\a G(z,z')  \, T_\n \, \cW'_\a \, 
\cD'^2 {\bar \cD}'^2 G(z',z) \Big) ~, 
\label{A2} \\
0 &=& \int {\rm d}^8 z   \int {\rm d}^8 z' \, 
G^{\m \n}(z,z') \,{\rm tr}_{\rm Ad}  \,\Big( T_\m \,
\cD^\a G(z,z')  \, \cW'_\a \, T_\n \,
\cD'^2 {\bar \cD}'^2 G(z',z) \label{A3} \\
&+& 2 T_\m\,  \cD^2 G(z,z')  \, \cW'^\a  \, T_\n \,  
\cD'_\a {\bar \cD}'^2 G(z',z)  
- T_\m \, \cD^2 G(z,z') \,  T_\n \, \cW'^\a  \, 
\cD'_\a {\bar \cD}'^2 G(z',z)  \Big)~, 
\non
\eea
which hold for the background  chosen.
The proofs of these identities are given in Appendix C.

The contribution (\ref{vector-fish-1.a}) requires more 
work to simplify. Let us first note the 
algebraic symmetry property 
\be 
\Big( G(z,z') \Big) ^{\rm T} = G(z',z)~,
\ee
and the differential identities \cite{KM2}
\be
{\bar \cD}^2 \, G(z,z') = {\bar \cD}'^2 \, G(z,z') ~, \qquad 
\cD^2 \, G(z,z') = \cD'^2 \, G(z,z')~.
\label{transpar}
\ee
Taken together, they imply 
\be
\int {\rm d}^8 z   \int {\rm d}^8 z' \, 
G^{\m \n}(z,z') \,{\rm tr}_{\rm Ad}
\Big( T_\m  \, [ {\bar \cD}^2 , \cD^2 ]
 G(z,z') \,T_\n \,
\{   {\bar \cD}'^2 , \cD'^2\} G(z',z) \Big) =0~.
\label{fancy}
\ee
Therefore,  one can replace
$$
 \cD^2 \, {\bar \cD}^2 \quad \longrightarrow 
\quad
\hf \{ \cD^2 , {\bar \cD}^2 \}
$$
in the third term on the right of 
(\ref{vector-fish-1.a}).
The number of spinor covariant derivatives 
in the first and fourth terms on the right 
of  (\ref{vector-fish-1.a}) can be reduced  
by making use of the identity 
(which holds for the background chosen)
\be 
{1 \over 16} {\bar \cD}^2 \,  \cD^2 \, {\bar \cD}^2 
=\Box_+ \, {\bar \cD}^2 
= {\bar \cD}^2 \, \Box_{\rm v} 
= \Box_{\rm v }\,  {\bar \cD}^2 
\ee
along with the equation 
\be
\Big( \Box_{\rm v} - \F \, \F^\dagger \Big) 
G(z,z') = - {\bf 1} \, \d^8(z-z')
\ee
which the Green's function under consideration 
obeys. Those contributions, in which a Green's function
 turns into the  delta-function  $\d^8(z-z')$, are no longer `fish' 
supergraphs; rather,  they become  `eight' diagrams.
Finally, it turns out that the second term on the right 
of  (\ref{vector-fish-1.a}) is identically zero, 
\bea
\int {\rm d}^8 z   \int {\rm d}^8 z' \, 
G^{\m \n}(z,z') \,{\rm tr}_{\rm Ad} \Big(
T_\m \,
 [ {\bar \cD}^2 , \cD^2 ] G(z,z') \, 
T_\n \, \cD'^\a {\bar \cD}'^2 \cD'_\a G(z',z) \Big)=0~,
\label{missed-zero}
\eea
for the background chosen.

The above considerations lead to
the following  expression  for $\D \G_{1}$:
\bea
\D \G_{1} &= & -  {1 \over 2^{6}} 
 \int {\rm d}^8 z   \int {\rm d}^8 z' \, 
G^{\m \n}(z,z') \,{\rm tr}_{\rm Ad} \Big(
{1 \over 2^5} T_\m
 \{ {\bar \cD}^2 ,  \cD^2 \}  G(z,z') 
\, T_\n \, \{ {\bar \cD}'^2 , \cD'^2 \}  G(z',z) \non \\
&+&  T_\m \,\cD^\a  G(z,z') \,  T_\n \, \F \F^\dagger  
\cD'_\a {\bar \cD}'^2 G(z',z) 
-   2 T_\m \,\cD^2   G(z,z') \,T_\n \,  \F \F^\dagger  
  {\bar \cD}'^2  G(z',z) \Big)  \non\\ 
&+& 
{1 \over 2^{6}}   \int {\rm d}^8 z  
\lim_{z' \to  z} \, {\bar \cD}^2 \Big\{
\cD^\a G^{\m \n}(z,z') \,{\rm tr}_{\rm Ad}
 \Big( T_\m \, \cD_\a G(z,z') \, T_\n \Big) \non \\
&& 
\phantom{ 
{1 \over 2^6 }   \int {\rm d}^8 z  
\lim_{z' \to  z} \, {\bar \cD}^2 }
~~ -
 G^{\m \n}(z,z') \,{\rm tr}_{\rm Ad}
 \Big( T_\m \, \cD^2 G(z,z') \, T_\n \Big) \Big\}~.
\eea
This may be further simplified
by noting the identity
\bea
&& ~~ \int {\rm d}^8 z   \int {\rm d}^8 z' \, 
G^{\m \n}(z,z') \,{\rm tr}_{\rm Ad} \Big(
T_\m \,\cD^\a  G(z,z') \,  T_\n \, \F \F^\dagger  
\cD'_\a {\bar \cD}'^2 G(z',z)  \Big) \non \\
&=& \hf 
 \int {\rm d}^8 z   \int {\rm d}^8 z' \, 
G^{\m \n}(z,z') \,{\rm tr}_{\rm Ad} \Big(
T_\m \,  G(z,z') \,  T_\n \, \F \F^\dagger  
\cD'^2 {\bar \cD}'^2 G(z',z)  \Big) ~.
\eea
The final expression for $\D \G_{1}$ is:
\bea
\D \G_{1} &= & -  {1 \over 2^{6}} 
 \int {\rm d}^8 z   \int {\rm d}^8 z' \, 
G^{\m \n}(z,z') \,{\rm tr}_{\rm Ad} \Big(
{1 \over 2^5} T_\m
 \{ {\bar \cD}^2 ,  \cD^2 \}  G(z,z') 
\, T_\n \, \{ {\bar \cD}'^2 , \cD'^2 \}  G(z',z) \non \\
&+&  \hf T_\m \,  G(z,z') \,  T_\n \, \F \F^\dagger  
\cD'^2 {\bar \cD}'^2 G(z',z) 
-   2 T_\m \,\cD^2   G(z,z') \,T_\n \,  \F \F^\dagger  
  {\bar \cD}'^2  G(z',z) \Big)  \non\\ 
&+& 
{1 \over 2^{6}}   \int {\rm d}^8 z  
\lim_{z' \to  z} \, {\bar \cD}^2 \Big\{
\cD^\a G^{\m \n}(z,z') \,{\rm tr}_{\rm Ad}
 \Big( T_\m \, \cD_\a G(z,z') \, T_\n \Big) \non \\
&& 
\phantom{ 
{1 \over 2^6 }   \int {\rm d}^8 z  
\lim_{z' \to  z} \, {\bar \cD}^2 }
~~ -
 G^{\m \n}(z,z') \,{\rm tr}_{\rm Ad}
 \Big( T_\m \, \cD^2 G(z,z') \, T_\n \Big) \Big\}~.
\label{vector-fish-1.b}
\eea

\subsection{Vector `eight' supergraphs}
The quartic interaction  (\ref{vect-4})
generates a number of `eight' supergraphs, 
\be 
\Big\langle S^{(4)}_{\rm vect}   
\Big\rangle_{\rm 1PI}  
\equiv \D \G_{3} ~.
\ee 
Their total contribution 
to the effective action is
\bea
 \D \G_{3} &= & 
{1 \over 2^4}  \int {\rm d}^8 z  
\lim_{z' \to  z} \,  \Big\{
\cD^\a G^{\m \n}(z,z') \,{\rm tr}_{\rm Ad}
 \Big( T_\m \, {\bar \cD}_\ad \cD_\a {\bar \cD}'^\ad 
G(z,z') \, T_\n \Big) \non \\
&& 
\phantom{+ {1 \over 2^4}  \int {\rm d}^8 z  
\lim_{z' \to  z} \,  } ~
-{1 \over 3}  G^{\m \n}(z,z') \,{\rm tr}_{\rm Ad}
 \Big( T_\m \, {\bar \cD}^2 \cD^2  G(z,z') \, T_\n \Big) 
\Big\} \non \\
&&- 
{1 \over 2^{6}}   \int {\rm d}^8 z  
\lim_{z' \to  z} \, {\bar \cD}^2 \Big\{
\cD^\a G^{\m \n}(z,z') \,{\rm tr}_{\rm Ad}
 \Big( T_\m \, \cD_\a G(z,z') \, T_\n \Big)  \non \\
&& 
\phantom{ 
{1 \over 2^6 }   \int {\rm d}^8 z  
\lim_{z' \to  z} \, {\bar \cD}^2 }
~~~~ -
 G^{\m \n}(z,z') \,{\rm tr}_{\rm Ad}
 \Big( T_\m \, \cD^2 G(z,z') \, T_\n \Big) \Big\}  ~.
\label{vector-eight}
\eea
The contributions
in the third and fourth lines of 
(\ref{vector-fish-1.b}) and 
(\ref{vector-eight}) cancel each other.

\subsection{Scalar `fish' supergraphs}
The cubic interaction  (\ref{scal-3})
generates the following 
fish-type contributions 
to the effective action 
\bea
{ {\rm i} \over 2}  \, \Big\langle S^{(3)}_{\rm scal} \, 
 S^{(3)}_{\rm scal} \Big\rangle_{\rm 1PI}  
\equiv \D \G_4~.
\eea
A direct evaluation gives
\bea
\D \G_4 &= &  {1 \over 2^{6}} 
 \int {\rm d}^8 z   \int {\rm d}^8 z' \, 
G^{\m \n}(z,z') \,{\rm tr}_{\rm Ad} 
\Big( {1 \over 8} T_\m
 \cD^2  {\bar \cD}^2   G(z,z') 
\, T_\n \, \cD'^2 {\bar \cD}'^2   G(z',z) \non \\
&+&  T_\m \F   \cD^2 {\bar \cD}^2   G(z,z') \,
T_\n \, \F^\dagger \,G(z',z)  
-2  T_\m \cD^2  {\bar \cD}^2   G(z,z') \,
T_\n \, \F^\dagger \,G(z',z) \,\F \Big)~.
\eea
With the use of (\ref{fancy}), 
we can transform $\D \G_4$ to the form
\bea
\D \G_4 &= &   {1 \over 2^{6}} 
 \int {\rm d}^8 z   \int {\rm d}^8 z' \, 
G^{\m \n}(z,z') \,{\rm tr}_{\rm Ad} 
\Big( {1 \over 2^5} T_\m \,
 \{ {\bar \cD}^2 , \cD^2 \}  G(z,z') 
\, T_\n \, \{ {\bar \cD}'^2 , \cD'^2 \}   G(z',z) \non \\
&& +
{1 \over 2^5} T_\m \,
 [ {\bar \cD}^2 , \cD^2 ]  G(z,z') 
\, T_\n \, [ {\bar \cD}'^2 , \cD'^2 ]   G(z',z)  
\label{scalar-fish} \\
&&+  T_\m \, \F \,  \cD^2  {\bar \cD}^2   G(z,z') \,
T_\n \, \F^\dagger \,G(z',z)  
-2 T_\m \, \cD^2  {\bar \cD}^2   G(z,z') \,
T_\n \, \F^\dagger \,G(z',z) \,\F \Big)~. \non 
\eea
As can be seen, the contributions in the first lines 
of  (\ref{vector-fish-1.b}) and (\ref{scalar-fish}) 
cancel each other.

\subsection{Scalar `eight' supergraphs}
The two-loop contribution to the effective action 
from the quartic interaction  (\ref{scal-4})
is extremely simple:
\bea
\D \G_5  \equiv  
\Big\langle S^{(4)}_{\rm scal}   
\Big\rangle_{\rm 1PI} 
={1 \over 2^{5}}   \int {\rm d}^8 z  
\lim_{z' \to  z} \,
G^{\m \n}(z,z') \,{\rm tr}_{\rm Ad}
 \Big( T_\m \, {\bar \cD}^2\cD^2 G(z,z') \, T_\n \Big) ~.
\label{scal-eight}
\eea

\subsection{Vector-scalar cross `fish' supergraphs}
A fish-type contribution
to the effective action,
\be
 {\rm i}  \, \Big\langle S^{(3)}_{\rm vect} \, 
 S^{(3)}_{\rm scal} \Big\rangle_{\rm 1PI}  
\equiv \D \G_6~,
\ee
 is generated by both 
the cubic vector and scalar interactions
(\ref{vect-3}) and  (\ref{scal-3}).
Its direct evaluation leads to
\bea
\D \G_6 &=& 
 {1 \over 2^{5}} 
 \int {\rm d}^8 z   \int {\rm d}^8 z' \, 
\cD^\a G^{\m \n}(z,z') \,{\rm tr}_{\rm Ad} 
\Big( T_\m \, {\bar \cD}^2   G(z,z') 
\, X_\n \, \cD'_\a   G(z',z) \Big) \non \\
&-& {1 \over 2^{5}} 
 \int {\rm d}^8 z   \int {\rm d}^8 z' \, 
 G^{\m \n}(z,z') \,{\rm tr}_{\rm Ad} 
\Big( T_\m \, {\bar \cD}^2   G(z,z') 
\, X_\n \, \cD'^2   G(z',z) \Big) ~,
\eea
where
\be
X_\n = \F \, T_\n \, \F^\dagger 
- \F^\dagger \, T_\n \, \F~.
\ee
It can be shown that the first term 
in $\D \G_6$ vanishes, and thus
\bea
\D \G_6 = - {1 \over 2^{5}} 
 \int {\rm d}^8 z   \int {\rm d}^8 z' \, 
 G^{\m \n}(z,z') \,{\rm tr}_{\rm Ad} 
\Big( T_\m \, {\bar \cD}^2   G(z,z') 
\, X_\n \, \cD'^2   G(z',z) \Big) ~.
\label{cross}
\eea

\subsection{Ghost  `fish' supergraphs}
The next step is to collect the 
two-loop fish-type supergraphs
generated by the cubic ghost action (\ref{gh-3}), 
\bea
{ {\rm i} \over 2}  \, \Big\langle S^{(3)}_{\rm gh} \, 
 S^{(3)}_{\rm gh} \Big\rangle_{\rm 1PI}  
\equiv \D \G_7~.
\eea
It is easy to recognize  that the second line 
in (\ref{gh-3}) does not contribute to $\D \G_7$. 
With this observation at our disposal, 
the remaining calculation is rather simple. 
The results is
\bea
\D \G_7 &= & -  {1 \over 2^{11}} 
 \int {\rm d}^8 z   \int {\rm d}^8 z' \, 
G^{\m \n}(z,z') \, \non \\
& & \qquad \qquad \times 
{\rm tr}_{\rm Ad} \Big(
 T_\m \, [ {\bar \cD}^2 ,  \cD^2 ]  G(z,z') 
\, T_\n \, [ {\bar \cD}'^2 , \cD'^2 ]  G(z',z) \Big)~.
\label{gh-fish}
\eea
It can be seen that $\D \G_7$ cancels 
the contribution in the second line of 
(\ref{scalar-fish}). 

\subsection{Ghost  `eight' supergraphs}
The two-loop contribution to the effective action 
from the quartic interaction  (\ref{gh-4})
is extremely simple:
\bea
\D \G_8  \equiv  
\Big\langle S^{(4)}_{\rm gh}   
\Big\rangle_{\rm 1PI} 
=- {1 \over 3 \cdot 2^5 }   \int {\rm d}^8 z  
\lim_{z' \to  z} \,
G^{\m \n}(z,z') \,{\rm tr}_{\rm Ad}
 \Big( T_\m \, {\bar \cD}^2\cD^2 G(z,z') \, T_\n \Big) ~.
\label{gh-eight}
\eea

\subsection{The pure \mbox{$\cN=2$} super 
Yang-Mills sector}
In this subsection, we would like 
to present 
the complete two-loop contribution to the effective 
action from the pure $\cN=2$  super Yang-Mills sector, 
including the ghosts. 
Several additional simplifications occur
in the case of the background (\ref{actual-background})
 we are actually interested in. 
${}$First of all,  it can be seen that the quantum 
correction (\ref{cross}) vanishes for 
the background (\ref{actual-background}), 
\be
\D \G_6 = 0~.
\ee
Second, the Green's function  commutes
with $\F$, $\F^\dagger$  and $\cW_\a$
(but not with ${\bar \cW}_\ad$)
for  the background chosen, 
\be
[\F , G(z,z') ] =[ \F^\dagger , G(z,z') ]=0~, 
\quad [\cW_\a , G(z,z')]=0~.
\ee
One can now prove the following identity
\bea
&&~~  \int {\rm d}^8 z   \int {\rm d}^8 z' \, 
G^{\m \n}(z,z') \,{\rm tr}_{\rm Ad} 
\Big( T_\m \F {\bar \cD}^2  \cD^2 G(z,z') 
\, T_\n G(z',z) \F^\dagger \Big)  \non \\
&=& \hf \int {\rm d}^8 z   \int {\rm d}^8 z' \, 
G^{\m \n}(z,z') \,{\rm tr}_{\rm Ad} 
\Big( T_\m \F^\dagger \F {\bar \cD}^2  \cD^2 G(z,z') 
\, T_\n G(z',z)  \Big) ~. 
\eea
This identity implies that 
the first term in the second line of 
(\ref{vector-fish-1.b})  cancels
the first term in the third line of
(\ref{scalar-fish}).

The complete two-loop contribution to the effective 
action from the pure $\cN=2$  super Yang-Mills sector, 
including the ghosts, is
\bea
\G_{\rm SYM} &=& 
\G_{\rm IV}  + \G_{\rm V} 
+ \G_{\rm VI} +\G_{\rm VII} + \G_{\rm VIII} ~,
\label{sym-two}
\eea
where 
\bea
 \G_{\rm IV} &=& 
{1 \over 2^6} 
 \int {\rm d}^8 z   \int {\rm d}^8 z' \, 
G^{\m \n}(z,z') \,{\rm tr}_{\rm Ad} 
\Big(  T_\m\, \cD^2 G(z,z') \,  T_\n \, \cW'^\a  \, 
\cD'_\a {\bar \cD}'^2 G(z',z) \Big)~, \non \\
\G_{\rm V} &=& 
- {1 \over 2^6} 
 \int {\rm d}^8 z   \int {\rm d}^8 z' \, 
G^{\m \n}(z,z') \,{\rm tr}_{\rm Ad} 
\Big(
T_\m\,  \cD^\a G(z,z')  \, \cW'_\a \, T_\n \,
\cD'^2 {\bar \cD}'^2 G(z',z) \Big) ~, \non \\
\G_{\rm VI} &=& 
{1 \over 2^5} 
\int {\rm d}^8 z   \int {\rm d}^8 z' \, 
G^{\m \n}(z,z') \,{\rm tr}_{\rm Ad} 
\Big(
T_\m \,\cD^2   G(z,z') \,T_\n \,  \F \F^\dagger  
  {\bar \cD}'^2  G(z',z) \Big)  ~, \non \\
\G_{\rm VII} &=& 
-{1 \over 2^5} 
\int {\rm d}^8 z   \int {\rm d}^8 z' \, 
G^{\m \n}(z,z') \,{\rm tr}_{\rm Ad} 
\Big(
T_\m \, \cD^2  {\bar \cD}^2   G(z,z') \,
T_\n \, \F^\dagger \,G(z',z) \,\F \Big)~, \non \\
\G_{\rm VIII} &= & 
{1 \over 2^4}  \int {\rm d}^8 z  
\lim_{z' \to  z} \,  
\cD^\a G^{\m \n}(z,z') \,{\rm tr}_{\rm Ad}
 \Big( T_\m \, {\bar \cD}_\ad \cD_\a {\bar \cD}'^\ad 
G(z,z') \, T_\n \Big) ~.
\eea
As follows from the relations
(\ref{vector-eight}), 
(\ref{scal-eight}) and (\ref{gh-eight}), 
all `eight' supergraphs of the form 
\bea
  \int {\rm d}^8 z  
\lim_{z' \to  z} \,
G^{\m \n}(z,z') \,{\rm tr}_{\rm Ad}
 \Big( T_\m \, {\bar \cD}^2\cD^2 G(z,z') \, T_\n \Big) 
\eea
cancel each other.

\sect{Specification of the  background  and 
related group-theoretical results}

At this stage, it is necessary to  describe 
the $SU(N)$ conventions adopted in the paper.
Lower-case Latin letters from the middle of the alphabet, 
$i,j,\dots$, 
will be used to denote matrix elements in the fundamental, 
with the convention $i =0,1,\dots, N-1 \equiv 0, \un{ i}$.
We choose a Cartan-Weyl basis 
to consist of the elements:
\be 
H_I = \{ H_0, H_{\un{I}}\}~, \quad  \un{I} = 1,\dots, N-2~, 
\qquad \quad E_{ij}~, \quad i\neq j~. 
\label{C-W}
\ee 
The basis elements in the fundamental representation 
are defined similarly to \cite{Georgi}, 
\bea
(E_{ij})_{kl} &=& \d_{ik}\, \d_{jl}~, \non \\
(H_I)_{kl} &=& \frac{1}{\sqrt{(N-I)(N-I-1)} }
\Big\{ (N-I)\, \d_{kI} \, \d_{lI} - 
\sum\limits_{i=I}^{N-1} \d_{ki} \, \d_{li} \Big\} ~,
\label{C-W-2}
\eea
and are characterized by the properties
\be
{\rm tr} (H_I\,H_J) = \d_{IJ}~, 
\qquad 
{\rm tr} (E_{ij}\,E_{kl}) = \d_{il}\,\d_{jk}~, 
\qquad {\rm tr} (H_I \,E_{kl}) =0~.
\ee
A generic element of the Lie algebra $su(N)$ is 
\be
v = v^I \, H_I + v^{ij} \, E_{ij} \equiv v^\m \,T_\m~,  
\qquad i \neq j ~, 
\label{generic}
\ee
${}$For $SU(N)$, the operation of trace 
in the adjoint representation, 
${\rm tr}_{\rm Ad}$, 
is related to that in the fundamental, 
${\rm tr}$, as follows
\be 
{\rm tr}_{\rm Ad} \,v^2 = 2N \, {\rm tr} \, v^2 ~,
\qquad  v \in su(N)~.
\ee 

The $\cN=2$ background vector multiplet is chosen to be
\be
\F = \f \, H_0~, \qquad \cW_\a = W_\a \, H_0~,
\label{actual-background}
\ee
Its characteristic feature is that it leaves
the subgroup $U(1) \times SU(N-1) \subset SU(N)$ 
unbroken, where $U(1)$ is associated with $H_0$
and  $SU(N-1)$ is  generated by 
$\{ H_{\un{I}}, \, E_{ \un{i} \un{j}} \}$.
In what follows, it is assumed that 
the gauge freedom associated with the broken
generators has been  used to bring 
the superfield connection 
$\G_A = \G^\m_A (z) T_\m$ in (\ref{gcd})
to the form $\G_A = \G^0_A (z) H_0$. 

The mass matrix is 
\be
|\cM|^2 = {\bar \f}\f \, (H_0)^2~, 
\ee
and therefore a superfield's mass is determined 
by  its $U(1)$ charge with respect to $H_0$. 
With the notation 
\be
e = \sqrt{ N /( N -1)}  ~,
\ee
the $U(1)$ charges of all quantum superfields 
are given in the table.
\begin{center}
\begin{tabular}{ | c || c | c | c | c |}  \hline
superfield &  
$v^{0\, \un{i} }$ & $v^{\un{i} \, 0}$ & 
$v^I$ & $v^{ \un{i} \,\un{j} }$\\ \hline 
$U(1) $ charge & 
$e $ & $-e $ 
&0&0\\  \hline
\end{tabular} \\
${}$\\
Table  1: $U(1)$ charges of superfields
\end{center}
Among the quantum gauge superfields, eq.
(\ref{generic}), 
there are $2(N-1)$ massive superfields 
($v^{0 \un{i}}$ and their conjugates $v^{ \un{i}0}$)
coupled to the background, 
while the remaining  $(N-1)^2$ superfields ($v^I$ and 
$v^{\un{i} \,\un{j}} $)  
do not interact with  the background 
and, therefore, are free massless.
This follows from the identity 
\be
[H_0 , E_{ij}] ~=~ \sqrt{N \over N-1}\,  \Big( 
\d_{0i}\, E_{0j} - \d_{0j}\, E_{i0} \Big)~. 
\ee

Since the basis 
(\ref{C-W}) is not orthonormal, 
${\rm tr}_{\rm F} (T_\m \, T_\n)  =g_{\m \n} \neq \d_{\m \n}$, 
it is now
necessary to keep track of the Cartan-Killing metric when 
working with adjoint vectors. For any elements
$u=u^\m T_\m $ and $v=v^\m T_\m $ of the Lie algebra, 
we have $u\cdot v =  {\rm tr}_{\rm F} (u\,v) =u^\m \,v_\m$, 
where $v_\m = g_{\m\n} v^\n$
($v_I =v^I$, $v_{ij}= v^{ji}$).

${}$For the background chosen, 
the Green's function
$\cG =(\cG^\m{}_\n)$, with 
\be
\cG^\m{}_\n (z,z') = G^{\m \l}(z,z') \, g_{\l \n} ~, 
\qquad  G^{\m \n}(z,z') = -{\rm i} 
\langle v^\m (z) \, v^\n (z') \rangle~,
\label{G}
\ee
is diagonal. Relative to the basis 
$T_\m = (H_I, E_{0 \un{i} }, E_{\un{i} 0} , E_{ \un{i} \un{j} })$, 
this Green's function has the form
\be
\cG = {\rm diag} \Big(\bG^{(0)}\, {\bf 1}_{N-1} , ~
\bG^{(e )}\, {\bf 1}_{N-1}, ~
\bG^{(-e )}\, {\bf 1}_{N-1}, ~
\bG^{(0)}\, {\bf 1}_{(N-1)(N-2)} \Big)~.
\label{I+II-ad12}
\ee
Here $\bG^{(e)} (z,z') $ denotes a $U(1)$
Green's function of charge $e$ under the 
equation 
\be
\Big(  \cD^a \cD_a - e\,W^\a \cD_\a 
+e \,{\bar W}_\ad {\bar \cD}^\ad -m^2 \Big) 
\bG^{(e)} (z,z') = -\d^8(z-z') ~,
\qquad m^2 = e^2 \,{\bar \f}\f ~,
\ee
with $\cD_A= D_A + {\rm i} \,e\, \G^0_A (z) $ 
the $U(1)$ gauge covariant derivatives.
Apart from the specific choice of a gauge group, 
its representation  and a mass term, 
this equation coincides with  (\ref{green}).
The heat kernel associated with $\bG^{(e)} (z,z') $
will be denoted $\bK^{(e)} (z,z'|s) $.

It follows from the analysis in section 4 that  all
the two-loop `fish' supergraphs have the 
following general form 
\bea
 \G_\ominus &= &
 \int {\rm d}^8 z   \int {\rm d}^8 z' \, 
G^{\m \n}(z,z') \,{\rm tr}_{\rm Ad} \,    
\Big( T_\m \,  \hat{\cG}(z,z') \, T_\n \,
\check{\cG}(z',z) \Big)~, 
\label{Gfish}
\eea
where
\be
 \hat{\cG}(z,z')  = A\, \cG(z,z') \, A' \equiv 
\hat{\cG}~, 
\qquad  
\check{\cG}(z',z) =B\, \cG(z,z') \, B' \equiv
\check{\cG}' ~,
\ee
with $A,\,A'$ and $B, \,B'$ some 
diagonal operators,   with respect to their 
$SU(N)$  indices, of the form
\be
A = {\rm diag} \Big(A^{(0)}\, {\bf 1}_{N-1} , ~
A^{(e )}\, {\bf 1}_{N-1}, ~
A^{(-e )}\, {\bf 1}_{N-1}, ~
A^{(0)}\, {\bf 1}_{(N-1)(N-2)} \Big)~,
\ee
and so on. We thus have
\be
\hat{\cG} = {\rm diag} \Big(\hat{\bG}^{(0)}\, {\bf 1}_{N-1} , ~
\hat{\bG}^{(e )}\, {\bf 1}_{N-1}, ~
\hat{\bG}^{(-e )}\, {\bf 1}_{N-1}, ~
\hat{\bG}^{(0)}\, {\bf 1}_{(N-1)(N-2)} \Big)~,
\label{hatG}
\ee
and similarly for $\check{\cG}' $. In 
the `fish' supergraphs listed in the previous
section,  the operators $A$ and $B$
are either the covariant  derivatives 
or some of the background
superfields $\F$, $\cW_\a $ and their conjugates, 
and products of these.

Using the results of \cite{KM3}, 
one obtains
\bea
{ \G_\ominus \over N(N-1)}
&= & 
2(N-2) \int {\rm d}^8 z   \int {\rm d}^8 z' \, 
\bG^{(0)}(z,z')  \hat{\bG}^{(0 )}
\check{\bG}'^{(0) } \non \\
&+& \int {\rm d}^8 z   \int {\rm d}^8 z' \, 
\bG^{(0)}(z,z') \left\{ \hat{\bG}^{(e )}
\check{\bG}'^{(e) }
+ \hat{\bG}^{(-e )}
\check{\bG}'^{(-e ) }
\right\}  
\non \\
&+& 
\int {\rm d}^8 z   \int {\rm d}^8 z' \, 
\left( 
\bG^{(e )}(z,z') \left\{ 
\hat{\bG}^{(-e )}
\check{\bG}'^{(0) }
+ \hat{\bG}^{(0 )}
\check{\bG}'^{(e) }
\right\} 
\right.  
\label{fish-structure-0}\\
&& \quad+ 
\left.
\bG^{(-e )}(z,z') \left\{ 
\hat{\bG}^{(e )}
\check{\bG}'^{(0) }
+ \hat{\bG}^{(0 )}
\check{\bG}'^{(-e) }
\right\}  \right)
~, \non
\eea
see Appendix D for the derivation.
${}$For the `fish' supergraphs under consideration, 
the expression in the first line is 
background-independent, and therefore 
\bea
{ \G_\ominus \over N(N-1)}
&= & 
\int {\rm d}^8 z   \int {\rm d}^8 z' \, 
\bG^{(0)}(z,z') \left\{ \hat{\bG}^{(e )}
\check{\bG}'^{(e) }
+ \hat{\bG}^{(-e )}
\check{\bG}'^{(-e ) }
\right\}  
\non \\
&+& 
\int {\rm d}^8 z   \int {\rm d}^8 z' \, 
\left( 
\bG^{(e )}(z,z') \left\{ 
\hat{\bG}^{(-e )}
\check{\bG}'^{(0) }
+ \hat{\bG}^{(0 )}
\check{\bG}'^{(e) }
\right\} 
\right.  
\label{fish-structure}\\
&& \quad+ 
\left.
\bG^{(-e )}(z,z') \left\{ 
\hat{\bG}^{(e )}
\check{\bG}'^{(0) }
+ \hat{\bG}^{(0 )}
\check{\bG}'^{(-e) }
\right\}  \right)
\equiv
 \G^{(0)} +   \G^{(e )} 
~. \non
\eea

The generic contribution to the effective action 
from an `eight' supergraph is of the form 
\bea
\G_\infty =  \int {\rm d}^8 z  
\lim_{z' \to  z} \,
\hat{G}^{\m \n}(z,z') \,{\rm tr}_{\rm Ad}
 \Big( T_\m T_\n \, \check{\cG}(z,z')  \Big) ~,
\qquad \hat{G}^{\m \n} = \hat{\cG}^\m{}_\l \,g^{\l \n}~,
\label{Geight}
\eea
where $\hat{\cG}$ and $\check{\cG}$ are of the type 
discussed above.
Using the group theoretic results obtained 
in \cite{KM3} and letting $G|=G(z,z)$, 
we get
\bea
\G_\infty &=& 2N(N-1)(N-2) 
 \int {\rm d}^8 z  \,\hat{\bG}^{(0)}| \,\check{\bG}^{(0)}|  
+ N(N-1) \int {\rm d}^8 z  \, \left\{
\hat{\bG}^{(0)}|  \Big( \check{\bG}^{(e)}| +\check{\bG}^{(-e)}| \Big)
\right. \non \\
 && \qquad  \qquad + \left.
 \hat{\bG}^{(e)}| 
\Big( \check{\bG}^{(0)}| +\check{\bG}^{(e)}| \Big)
+\hat{\bG}^{(-e)} |
\Big( \check{\bG}^{(0)}| +\check{\bG}^{(-e)}| \Big)
\right\}~,
\label{eight-structure-0}
\eea
see Appendix D for the derivation.
${}$For the `eight' supergraphs under consideration, 
either $\hat{\bG}^{(0)}| =0$ or $\check{\bG}^{(0)}|  =0$, 
and therefore
\bea
\G_\infty 
&=&  
N(N-1) 
\int {\rm d}^8 z  \, \left\{
\hat{\bG}^{(0)}|  
\Big( \check{\bG}^{(e)}| +\check{\bG}^{(-e)}| \Big)
\right. \non \\
 && \qquad  \qquad + \left.
 \hat{\bG}^{(e)}| 
\Big( \check{\bG}^{(0)}| +\check{\bG}^{(e)}| \Big)
+\hat{\bG}^{(-e)} |
\Big( \check{\bG}^{(0)}| +\check{\bG}^{(-e)}| \Big)
\right\}~.
\label{eight-structure}
\eea
As follows from (\ref{fish-structure}) 
and (\ref{eight-structure}), the two-loop effective 
action contains a common factor $N(N-1)$.

\sect{The hypermultiplet sector}

In this section, we focus on evaluating 
the two-loop contributions to the effective action 
from the hypermultiplet sector.

\subsection{Evaluation of \mbox{$ \G_{\rm I} $}}

The hardest quantum correction to compute is
\bea
 \G_{\rm I} &= &   {1 \over 2^{9}} 
 \int {\rm d}^8 z   \int {\rm d}^8 z' \, 
G^{\m \n}(z,z') 
{\rm tr}_{\rm Ad} \Big(
 T_\m \, [ {\bar \cD}^2 ,  \cD^2 ]  \cG(z,z') 
\, T_\n \, [ {\bar \cD}'^2 , \cD'^2 ]  \cG(z',z) \Big)~.
\eea
${}$Following the notation introduced at
the end of the previous section, we now have
\be
\hat{\cG} = [ {\bar \cD}^2 ,  \cD^2 ]  \cG(z,z') ~,
\qquad
\check{\cG}' = [ {\bar \cD}'^2 , \cD'^2 ]  \cG(z',z) ~,
\ee
and the relevant $U(1)$  components
of $\hat{\cG}$ and $\check{\cG}' $ are 
\bea
\hat{\bG}^{(e)} &=& [ {\bar \cD}^2 ,  \cD^2 ]  
\bG^{(e)} (z,z') ~, \non \\
\check{\bG}'^{(e)} &=&
 [ {\bar \cD}'^2 , \cD'^2 ]  \bG^{(e)}(z',z) 
= - [ {\bar \cD}^2 ,  \cD^2 ]  \bG^{(e)}(z',z) 
= - [ {\bar \cD}^2 ,  \cD^2 ]  \bG^{(-e)}(z,z') ~.
\label{hat-check-1}
\eea

In computing $ \G_{\rm I}^{(0)}$
and $ \G_{\rm I}^{(e)}$, 
a key technical observation is the identity
\cite{KM2}
\bea
\frac{1}{16} [ {\bar \cD}^2, \cD^2 ] \, \bK^{(e)}(z,z' | s) 
& \approx  &
\frac{\rm i}{(4 \pi s)^2} \, 
\sqrt{ \det
\left( \frac{s \,eF}{\sinh (s\,e F)}\right) } \,
\Big( \r \, \frac{2e F}{{\rm e}^{ 2  s \,e F} -1} \Big)^{\a \ad} \,
\non \\
&\times&
\z^{(e)}_\a (s) \,  \bar{\z}^{(e)}_\ad (s)  \,
{\rm e}^{ \frac{{\rm i}}{4} 
\r \, eF \coth ( s \, e F) \, \r } \, I(z,z') ~,
\eea
where we have omitted all terms of at least third order 
in the Grassmann variables  $\z_\a, \,{\bar \z}_\ad$ and 
$W_\a, \, {\bar W}_\ad$ 
as they  do not contribute to $ \G_{\rm I}$.
In particular, it makes  the evaluation of one of the 
two Grassmann integrals trivial, say, the integral over  $\q'$.
The term $ \G_{\rm I}^{(0)}$ involves a Grassmann 
integral of the form
\bea
\int {\rm d}^4 \q'   \, \d^2 (\z  )\, 
\d^2( \bar{\z}) \, \z^{(e)}_\a (s) \,  \bar{\z}^{(e)}_\ad (s)  \,
\z^{(-e)}_\b (t) \,  \bar{\z}^{(-e)}_\bd (t)
= \left(  \z^{(e)}_\a (s) \,  \bar{\z}^{(e)}_\ad (s)  \,
\z^{(-e)}_\b (t) \,  \bar{\z}^{(-e)}_\bd (t) 
\right)\Big|_{\z ={\bar \z}=0} ~,\non 
\eea
where the Grassmann delta-fuction
$ \d^2 (\z  )\, \d^2( \bar{\z}) $ comes from the 
heat kernel corresponding to $\bG^{(0)}(z,z') $.
The expression obtained is proportional to 
$W^2 \,{\bar W}^2$.  The $ \G_{\rm I}^{(e )}$ 
involves a Grassmann integral of the form
\bea
\int {\rm d}^4 \q'   \, \d^2 (\z^{(e)}  (s))\, 
\d^2( \bar{\z}^{(e)}(s) ) \, \z_\a  \,  \bar{\z}_\ad   \,
\z^{(-e)}_\b (t) \,  \bar{\z}^{(-e)}_\bd (t)
= \left(  \z_\a  \,  \bar{\z}_\ad   \,
\z^{(-e)}_\b (t) \,  \bar{\z}^{(-e)}_\bd (t) 
\right)\Big| \Big|  ~,
\non 
\eea
where the symbol $| |$ indicates that one should set
\bea
\z^{(e)}_\a(s) ={\bar \z}^{(e)}_\ad(s)=0~, \non  
\eea
with $z^{(e)}(s)$ and ${\bar \z}^{(e)}(s)$ 
defined in eq. (\ref{shifted}).

The next step in evaluating $ \G_{\rm I}^{(0)}$
and $ \G_{\rm I}^{(e)}$
is to do one of the space-time integrals, say, 
the integral over $x'$. 
Replacing  the bosonic integration variables 
 by the rule $\{x, \, x' \}  \to \{x, \, \r\}$, 
 we end up with with a Gaussian integral 
of the form 
\bea
\frac{1}{(4\p)^2} \int {\rm d}^4 \r \, \r^2\,
{\rm e}^{  {\rm i}  \r^2  A   /4 } 
&=& - \frac{ 8 }{ A^3 } ~, 
\label{Gauss-1}
\eea
where 
\bea
A&=& {1 \over u}  +\U(s,t)~, 
\qquad 
\U(s,t) =
\frac{e \bar B }{2} \, 
\frac{ \sin (e {\bar B} (s+t) /2) } 
{\sin (s\,e{\bar B} /2) \sin (t\,e{\bar B} /2) }~.
\label{A}
\eea
Here we have used the explicit structure of the heat 
kernel,  described in Appendix B. 
The proper-time parameter $u$ in (\ref{A}) corresponds
to a free kernel, $\bK^{(0)}(z,z'|u) $, 
while the proper-time parameters $s$ and $t$ 
correspond to charged kernels, 
$\bK^{(e)}(z,z'|s) $ and $\bK^{(-e)}(z,z'|t) $.

Let us briefly describe the evaluation\footnote{The 
quantum correction
$ \G_{\rm I}^{(0)} $ actually coincides 
 with a special sector (denoted by $\G_{\rm I+II}$ in \cite{KM2})
of the two-loop effective action for $\cN=2$ SQED
computed in \cite{KM2}.} of $ \G_{\rm I}^{(0)} $.  
The two terms in 
\bea
 \G_{\rm I}^{(0)} &= &   {1 \over 2^{9}} 
 \int {\rm d}^8 z   \int {\rm d}^8 z' \, 
\bG^{(0)}(z,z') 
\left\{  \hat{\bG}^{(e )}
\check{\bG}'^{(e ) }
+ \hat{\bG}^{(-e )}
\check{\bG}'^{(-e ) }
\right\}  
\eea
turn out to produce the same  contribution,
 therefore
\bea
 \G_{\rm I}^{(0)} &= &   {1 \over 2^8 } 
 \int {\rm d}^8 z   \int {\rm d}^8 z' \, 
\bG^{(0)}(z,z') \,  \hat{\bG}^{(e )} \,
\check{\bG}'^{(e ) }~.
\eea

Direct calculations lead to 
\bea
 \G_{\rm I}^{(0)} &= &
-\frac{4e^4}{(4\p)^4 }
\int {\rm d}^8 z \, W^2 {\bar W}^2 
\int \limits_{0}^{\infty} 
\frac{ {\rm d}s  \,{\rm d}t \,{\rm d}u}
{(stu)^2} \, 
\frac{st}{[u^{-1} +\U(s,t)]^3} \,
{\rm e}^{-{\rm i} e^2({\bar \f}\f -{\rm i}\ve) (s+t)} \non \\
& \times & \L(s\,e{\bar B}/2) \, \L(t\,e{\bar B}/2)~,
\label{G-I-0}
\eea
with $\U(s,t)$ 
and $\L(s\,e{\bar B}/2)$ defined in (\ref{A})
and (\ref{lambda}), respectively.
The $u$-integral here is elementary. 
Making use of the explicit expressions for 
$\U(s,t)$
and $\L(s\,e{\bar B}/2)$, one obtains
\bea
 \G_{\rm I}^{(0)} &= &
-\frac{2e^4}{(4\p)^4 }
\int {\rm d}^8 z \, W^2 {\bar W}^2 
\int \limits_{0}^{\infty}  {\rm d}s 
\int \limits_{0}^{\infty}  {\rm d}t \,
\frac{ st \, (e { \bar B }/2)^2 } 
{ \sin^2 (e {\bar B} (s+t) /2) } 
{\rm e}^{-{\rm i} e^2({\bar \f}\f -{\rm i}\ve) (s+t)} ~.
\eea
In this paper, we are only interested in the real part of
the effective action, and therefore   
the Wick rotation $s \to -{\rm i} s$ and 
$t \to -{\rm i} t$ can be implemented naively. 
After re-scaling the proper-time
variables, one obtains 
\bea
 \G_{\rm I}^{(0)} &= &
\frac{2}{(4\p)^4 }
\int {\rm d}^8 z \, 
\frac{W^2 {\bar W}^2 }{(\f {\bar \f})^2}
\int \limits_{0}^{\infty}  {\rm d}s 
\int \limits_{0}^{\infty}  {\rm d}t \,
\frac{ st \, (\J/2e)^2 } 
{ \sinh^2 (\J (s+t) /2e) } 
{\rm e}^{- (s+t)} ~.
\label{G-I-1}
\eea

The double proper-time integral 
in (\ref{G-I-1})
can be reduced to a single integral, 
by introducing new integration variables,
$\a$ and  $\t $, defined as follows
(see, e.g.,   \cite{DR})
\bea 
s+t = \t~,  \quad  s-t =\t \,\a~, \quad
\qquad \t \in [0, \infty ) ~, 
\quad \a \in [-1, 1]~, 
\label{change-of-var}
\eea
such that 
\be
\int \limits_{0}^{\infty} {\rm d}s
\int \limits_{0}^{\infty}  {\rm d}t \, 
L\Big(s, t \Big ) = 
\hf \int \limits_{0}^{\infty} {\rm d}\t 
\int \limits_{-1}^{+1} {\rm d}\a \, \t \, 
L\Big( s(\a, \t), t(\a, \t) \Big) ~.
\ee
This leads to 
\bea
 \G_{\rm I}^{(0)} &= &
\frac{1}{3(4\p)^4 }
\int {\rm d}^8 z \, 
\frac{W^2 {\bar W}^2 }{(\f {\bar \f})^2} \non \\
&+&
\frac{1}{3(4\p)^4 }
\int {\rm d}^8 z \, 
\frac{W^2 {\bar W}^2 }{(\f {\bar \f})^2}
\int \limits_{0}^{\infty}  {\rm d}s \,s^3\,
\left\{ \frac{ ( \J/2e)^2 } 
{ \sinh^2 (s\,  \J  /2e) } -{1\over s^2} \right\}
{\rm e}^{- s} ~.
\label{G-I-2}
\eea

We now turn to 
\bea
 \G_{\rm I}^{(e)} &=&{1 \over 2^9 } 
\int {\rm d}^8 z   \int {\rm d}^8 z' \, 
\left( 
\bG^{(e )}(z,z') \left\{ 
\hat{\bG}^{(-e )}
\check{\bG}'^{(0) }
 +  \hat{\bG}^{(0 )}
\check{\bG}'^{(e) }
\right\} 
\right.   \non \\
&& \qquad \qquad  \qquad
+ 
\left.
\bG^{(-e )}(z,z') \left\{ 
\hat{\bG}^{(e )}
\check{\bG}'^{(0) }
+ \hat{\bG}^{(0 )}
\check{\bG}'^{(-e) }
\right\}  \right) ~,
\eea
with  $\hat{\bG}^{(e)}$ and 
$\check{\bG}'^{(e) }$ defined in 
(\ref{hat-check-1}). 
It can be shown that the four terms
in $ \G_{\rm I}^{(e)}$ produce identical
contributions, hence
\bea
 \G_{\rm I}^{(e)} 
& =&{1 \over 2^7 } 
\int {\rm d}^8 z   \int {\rm d}^8 z' \, 
\bG^{(e )}(z,z') 
\hat{\bG}^{(-e )}
\check{\bG}'^{(0)} ~.
\eea
Unlike $ \G_{\rm I}^{(0)}$,
no quantum corrections of 
this type occur in the 
case of $\cN=2$ SQED \cite{KM2}. 
Therefore,  the quantum correction $ \G_{\rm I}^{(e)}$ 
is non-abelian in origin.  

The evaluation of $ \G_{\rm I}^{(e)}$ follows the 
steps outlined above. The result is 
\bea
 \G_{\rm I}^{(e)} &= &
\frac{4e^4}{(4\p)^4 }
\int {\rm d}^8 z \, W^2 {\bar W}^2 
\int \limits_{0}^{\infty}  {\rm d}s 
\int \limits_{0}^{\infty}  {\rm d}t \,
\frac{ s(s+t) \, (e { \bar B }/2)^2 } 
{ \sinh^2 (t\,e {\bar B}  /2) } 
{\rm e}^{ -e^2 {\bar \f}\f  (s+t)} \non \\
&=&\frac{4}{(4\p)^4 }
\int {\rm d}^8 z \, \frac{W^2 {\bar W}^2}
{(\f {\bar \f})^2} 
 \int \limits_{0}^{\infty}  {\rm d}t \,
\frac{ (e { \bar B }/2)^2 } 
{ \sinh^2 (t\,e {\bar B}  /2) } 
\left\{ {2\over  e^2 {\bar \f}\f} +t \right\}
\,{\rm e}^{ -e^2 {\bar \f}\f t   } ~.
\eea
This quantum correction involves 
a divergent proper-time integral. 
Let us separate its finite and divergent pieces, 
\bea
 \G_{\rm I}^{(e)} &= &
\frac{4}{(4\p)^4 }
\int {\rm d}^8 z \, \frac{W^2 {\bar W}^2}
{(\f {\bar \f})^2} 
 \int \limits_{0}^{\infty}  {\rm d}s \,
(2 +s ) \left\{ \frac{ ( \J/2e)^2 } 
{ \sinh^2 (s\,  \J  /2e) } -{1\over s^2} \right\}
{\rm e}^{- s} \non \\
&+& 
\frac{4}{(4\p)^4 }
\int {\rm d}^8 z \, \frac{W^2 {\bar W}^2}
{(\f {\bar \f})^2} 
 \int \limits_{0}^{\infty}  {\rm d}t \,
\left\{ {2\over  e^2 {\bar \f}\f t^2} +{1 \over t}
 \right\}
\,{\rm e}^{ -e^2 {\bar \f}\f t   } ~.
\label{G-I-e}
\eea
The expression in the first line is finite,
while  that in the second line
contains a divergent proper-time integral.
We are going to show that 
its divergence is cancelled against 
the divergent part of $\D \G_{\rm III}$.

\subsection{Cancellation of divergences}
On the basis of direct calculations, one finds
\bea
{ \G_{\rm III} \over N(N-1) } 
&=& -
\frac{2e^4}{(4\p)^4 }
\int {\rm d}^8 z \, W^2 {\bar W}^2 
\int \limits_{0}^{\infty}  {\rm d}s \,s^2
\,{\rm e}^{ -e^2 {\bar \f}\f s  } 
\int \limits_{0}^{\infty}  {\rm d}t 
\left\{ \frac{ (e { \bar B }/2)^2 } 
{ \sinh^2 (t\,e {\bar B}  /2) } \,
{\rm e}^{ -e^2 {\bar \f}\f t  } 
+{1 \over t^2} \right\} \non \\
&=& -
\frac{4}{(4\p)^4 }
\int {\rm d}^8 z \, 
\frac{W^2 {\bar W}^2 }{(\f {\bar \f})^2}
\int \limits_{0}^{\infty}  {\rm d}s 
\left\{ \frac{ ( \J/2e)^2 } 
{ \sinh^2 (s\,  \J  /2e) } -{1\over s^2} \right\} 
\,{\rm e}^{ - s  }
\non \\
&&- 
\frac{4}{(4\p)^4 }
\int {\rm d}^8 z \, 
\frac{W^2 {\bar W}^2 }{(\f {\bar \f})^2}
\int \limits_{0}^{\infty} 
{  {\rm d} t \over e^2 \f {\bar \f}\, t^2}  
\left\{  {\rm e}^{ -e^2 \f {\bar \f} t  }  + 1 \right\}~.
\eea
Here the expression in the second line is 
finite, while the one in the third line contains
a divergent proper-time integral. 
Let us combine this divergent contribution 
with that in the second line of 
(\ref{G-I-e}):
\bea
&&
\int \limits_{0}^{\infty}  {\rm d}t \,
\left\{ {2\over  e^2 {\bar \f}\f \,t^2} +{1 \over t}
 \right\}\,{\rm e}^{ -e^2 {\bar \f}\f t   } 
-\int \limits_{0}^{\infty} 
{  {\rm d} t \over e^2 \f {\bar \f}\, t^2}  
\left\{  {\rm e}^{ -e^2 \f {\bar \f} t  }  + 1 \right\} \non \\
&& \qquad \qquad
=  \int \limits_{0}^{\infty}  {\rm d}t \,
{ {\rm d} \over  {\rm d}t  } 
\left\{ 
\frac{ 1-   {\rm e}^{ -e^2 \f {\bar \f } t  } }
{e^2 {\bar \f}\f \,t} 
\right\} =-1~.
\eea
This finite contribution will prove in Section 8
 to be vital for the absence of $F^4$ quantum corrections 
at two loops. 

\subsection{Evaluation of \mbox{$ \G_{\rm II} $}}
The remaining hypermultiplet quantum correction 
is
\bea
 \G_{\rm II}
=- { 1 \over 2^4} \int {\rm d}^8 z   \int {\rm d}^8 z' \, 
G^{\m \n}(z,z') {\rm tr}_{\rm Ad} \Big(
 T_\m \, \F^\dagger \, {\bar \cD}^2   \cG(z,z') 
\, T_\n \, \F \,   \cD'^2   \cG(z',z) \Big)~.
\eea
${}$Following the notation introduced at
the end of section 5, we  now have
\be
\hat{\cG} = \F^\dagger \,{\bar \cD}^2   \cG(z,z') ~,
\qquad
\check{\cG}' = \F \, \cD'^2   \cG(z',z) ~,
\ee
and the relevant $U(1)$  components
of $\hat{\cG}$ and $\check{\cG}' $ are 
\bea
\hat{\bG}^{(e)} = e{\bar \f}\, {\bar \cD}^2  
\bG^{(e)} (z,z') ~, \quad && \quad
\check{\bG}'^{(e)} =
 e\f \,\cD'^2   \bG^{(e)}(z',z) 
= e\f \, \cD^2   \bG^{(-e)}(z,z') ~, \non \\
\hat{\bG}^{(0)} &=& \check{\bG}'^{(0)} =0~. 
\label{hat-check-2}
\eea
We thus have 
\be
 \G_{\rm II}^{(e)}=0~, \qquad
 \G_{\rm II}^{(0)}
 =- { 1 \over 2^3} \int {\rm d}^8 z   \int {\rm d}^8 z' \, 
G^{\m \n}(z,z') \, \hat{\bG}^{(e)} 
\, \check{\bG}'^{(e)} ~.
\ee

As compared with the evaluation of 
$ \G_{\rm I}^{(0)}$, the procedure
for computing $ \G_{\rm II}^{(0)}$
differs in three points. First of all, 
the quantum correction $ \G_{\rm II}^{(0)}$ 
involves a Grassmann 
integral of the form
\bea
\int {\rm d}^4 \q'   \, \d^2 (\z  )\, 
\d^2( \bar{\z}) \, 
\d^2(\z^{(e)} (s)) \,  
\d^2( \bar{\z}^{(-e)} (t)) 
= \left(  
\d^2(\z^{(e)} (s)) \,  
\d^2( \bar{\z}^{(-e)} (t)) 
\right)
\Big|_{\z ={\bar \z}=0} ~.\non 
\eea
Second, instead of the Gaussian integral 
(\ref{Gauss-1}),  $ \G_{\rm II}^{(0)}$ 
involves the following integral 
\bea
\frac{1}{(4\p)^2} \int {\rm d}^4 \r \, \r^2\,
{\rm e}^{  {\rm i}  \r^2  A   /4 } 
&=&  \frac{ \rm i }{ A^2 } ~, 
\label{Gauss-2}
\eea
with $A$ defined in (\ref{A}).
Finally, in contrast to the $u$-integral
in   (\ref{G-I-0}), 
$ \G_{\rm II}^{(0)}$ 
involves the following proper-time integral
\be 
\int \limits_{0}^{\infty} 
\frac{ {\rm d}u}
{u^2} \, 
\frac{1}{[u^{-1} +\U(s,t)]^2} 
= {1 \over \U(s,t)}~.
\ee

Direct calculations lead to 
\bea
 \G_{\rm II}^{(0)} &= &
\frac{2e^4}{(4\p)^4 }
\int {\rm d}^8 z \, W^2 {\bar W}^2
\int \limits_{0}^{\infty}  {\rm d}s 
\int \limits_{0}^{\infty}  {\rm d}t \,
\frac{ s^2 \, e^2\f {\bar \f}\,
(e { \bar B }/2) \, \sinh (t\,e{\bar B}/2) } 
{ \sinh (s\,e{\bar B}/2)  
\sinh (e {\bar B} (s+t) /2) } 
{\rm e}^{- e^2{\bar \f}\f (s+t)} \non \\
&= &
\frac{2}{(4\p)^4 }
\int {\rm d}^8 z \, 
\frac{W^2 {\bar W}^2}{(\f {\bar \f})^2}
\int \limits_{0}^{\infty}  {\rm d}s 
\int \limits_{0}^{\infty}  {\rm d}t \,
\frac{ s^2  (\J/2e) \, \sinh (t\,\J/2e) } 
{ \sinh (s\,\J/2e)  
\sinh (\J (s+t) /2e) } 
{\rm e}^{-  (s+t)}~.
\eea
Using the identity
\be
\frac{ \sinh t } {\sinh s \, \sinh (s+t) }
= \coth s - \coth (s+t)~, 
\ee
we can rewrite $ \G_{\rm II}^{(0)}$
as follows 
\bea
 \G_{\rm II}^{(0)} &= &
\frac{2}{(4\p)^4 }
\int {\rm d}^8 z \, 
\frac{W^2 {\bar W}^2}{(\f {\bar \f})^2}
\int \limits_{0}^{\infty}  {\rm d}t \,{\rm e}^{-t } 
\int \limits_{0}^{\infty}  {\rm d}s \,s^2
(\J / 2e) \, \coth (s\J/2e) \, {\rm e}^{-s } \non \\
&-&
\frac{2}{(4\p)^4 }
\int {\rm d}^8 z \, 
\frac{W^2 {\bar W}^2}{(\f {\bar \f})^2}
\int \limits_{0}^{\infty}  {\rm d}s 
\int \limits_{0}^{\infty}  {\rm d}t \,s^2
(\J / 2e) \,  \coth ( \J (s+t)  /2e ) \,
{\rm e}^{-(s+t) }~. 
\eea
In the expression in the first line, 
the proper-time integrals are factorized, 
and one of them is elementary.  For
the expression in the second line, 
the corresponding double proper-time integral 
can be reduced to a single integral, 
by implementing the change of 
variables (\ref{change-of-var}).
We end up with 
\bea
 \G_{\rm II}^{(0)} &= &
\frac{2}{(4\p)^4 }
\int {\rm d}^8 z \, 
\frac{W^2 {\bar W}^2}{(\f {\bar \f})^2}
\int \limits_{0}^{\infty}  {\rm d}s \,
\Big( s^2 -{1\over 3}s^3\Big)\,
\left\{
(\J / 2e) \, \coth (s\J/2e) 
-{1\over s} \right\} \, {\rm e}^{-s } \non \\
&+& \frac{2}{3(4\p)^4 }
\int {\rm d}^8 z \, 
\frac{W^2 {\bar W}^2}{(\f {\bar \f})^2} ~. 
\label{G-II}
\eea

\subsection{Hypermultiplet effective action}
Now, as the quantum corrections
$\G_{\rm I}$, $\G_{\rm II}$ and $\G_{\rm III}$
have been computed, we can put these parts together
to obtain the total contribution, 
$\G_{\rm hyper} =
 \G_{\rm I}+\G_{\rm II}+\G_{\rm III}$, 
to the effective action from 
all the two-loop supergraphs generated by 
$S^{(3)}_{\rm hyper}$ and $S^{(4)}_{\rm hyper}$.
There still remains, however, 
one technical point to take care of.
Looking at (\ref{G-II}),  we see that
$ \G_{\rm II}$ involves the function 
\be
\left\{
(\J / 2e) \, \coth (s\J/2e) 
-{1\over s} \right\} ~,
\label{coth}
\ee
while both $ \G_{\rm I}$ and $ \G_{\rm III}$ 
involve a slightly different but related function
\be
 \left\{ 
\frac{ (\J / 2e) ^2 }  { \sinh^2 (s \J /2e) } 
-{1 \over s^2} \right\}  = - 
{ {\rm d} \over {\rm d} s} 
\left\{  (\J / 2e)\, { \coth (s \J /2e) } 
-{1 \over s} \right\}  ~.
\label{by-parts}
\ee
Using this identity, we can integrate by parts
all the proper-time integrals in $\G_{\rm I}+\G_{\rm III}$
in order to convert the function 
(\ref{by-parts}) into (\ref{coth}).
We thus end up with 
\bea
{\G_{\rm hyper} \over N(N-1)}
&=& 
- \frac{3}{(4\p)^4 }
\int {\rm d}^8 z \, 
\frac{W^2 {\bar W}^2}{(\f {\bar \f})^2}
\label{final-hyper} \\
&&-
\frac{1}{(4\p)^4 }
\int {\rm d}^8 z \, 
\frac{W^2 {\bar W}^2}{(\f {\bar \f})^2}
\int \limits_{0}^{\infty}  {\rm d}s \,
\Big( 4s -3 s^2 + s^3\Big)\,
\left\{
(\J / 2e)  \coth (s\J/2e) 
-{1\over s} \right\} \, {\rm e}^{-s } ~.\non
\eea
In the second term, the proper-time integral 
involves the cubic polynomial 
$\cP(s) = 4s -3 s^2 + s^3$ such that
$\cP(0) =0$. The latter property 
is simply a consequence of the fact 
that  there are no divergences.

\sect{The super Yang-Mills sector}

Evaluation of the supergraphs
$\G_{\rm IV}$, $\G_{\rm V}$, 
$\G_{\rm VI}$ and $\G_{\rm VII}$
is very similar to that of $\G_{\rm II}$
described in the previous section.
We therefore give only the final results:
\bea
\G_{\rm IV}^{(0)} &= &
-\frac{1}{(4\p)^4 }
\int {\rm d}^8 z \, 
\frac{W^2 {\bar W}^2}{(\f {\bar \f})^2}
\int \limits_{0}^{\infty}  {\rm d}s \,
\int \limits_{0}^{\infty}  {\rm d}t \, s\,
(\J / 2e) \, \coth (\J (s+t)/2e) \, {\rm e}^{-(s+t) } \non \\
&+& \frac{1}{(4\p)^4 }
\int {\rm d}^8 z \, 
\frac{W^2 {\bar W}^2}{(\f {\bar \f})^2}
\int \limits_{0}^{\infty}  {\rm d}s \,
s\, (\J / 2e) \, \coth (s \J /2e) \, {\rm e}^{-s } ~, \non \\
\G_{\rm IV}^{(e)} &= &
-\frac{1}{(4\p)^4 }
\int {\rm d}^8 z \, 
\frac{W^2 {\bar W}^2}{(\f {\bar \f})^2}
\int \limits_{0}^{\infty}  {\rm d}s \,
\int \limits_{0}^{\infty}  {\rm d}t \, (s+t)\,
(\J / 2e) \, \coth (\J (s+t)/2e) \, {\rm e}^{-(s+t) } \non \\
&+& \frac{1}{(4\p)^4 }
\int {\rm d}^8 z \, 
\frac{W^2 {\bar W}^2}{(\f {\bar \f})^2}
\int \limits_{0}^{\infty}  {\rm d}s \,
(1+s)\, (\J / 2e) \, \coth (s \J /2e) \, {\rm e}^{-s }~. 
\label{IV}\\
\G_{\rm V}^{(0)} &= &
\frac{1}{(4\p)^4 }
\int {\rm d}^8 z \, 
\frac{W^2 {\bar W}^2}{(\f {\bar \f})^2}
\int \limits_{0}^{\infty}  {\rm d}s \,
\int \limits_{0}^{\infty}  {\rm d}t \, s\,
(\J / 2e) \, \coth (\J (s+t)/2e) \, {\rm e}^{-(s+t) } \non \\
&-& \frac{1}{(4\p)^4 }
\int {\rm d}^8 z \, 
\frac{W^2 {\bar W}^2}{(\f {\bar \f})^2}
\int \limits_{0}^{\infty}  {\rm d}s \,
(\J / 2e) \, \coth (s \J /2e) \, {\rm e}^{-s } ~, \non \\
\G_{\rm V}^{(e)} &= &
-  \frac{2}{(4\p)^4 }
\int {\rm d}^8 z \, 
\frac{W^2 {\bar W}^2}{(\f {\bar \f})^2}
\int \limits_{0}^{\infty}  {\rm d}s \,(1+s)\,
(\J / 2e) \, \coth (s \J /2e) \, {\rm e}^{-s } ~, 
\label{V}\\
\G_{\rm VI}^{(0)} &= &
\frac{1}{(4\p)^4 }
\int {\rm d}^8 z \, 
\frac{W^2 {\bar W}^2}{(\f {\bar \f})^2}
\int \limits_{0}^{\infty}  {\rm d}s \,
\int \limits_{0}^{\infty}  {\rm d}t \, s^2\,
(\J / 2e) \, \coth (\J (s+t)/2e) \, {\rm e}^{-(s+t) } \non \\
&-& \frac{1}{(4\p)^4 }
\int {\rm d}^8 z \, 
\frac{W^2 {\bar W}^2}{(\f {\bar \f})^2}
\int \limits_{0}^{\infty}  {\rm d}s \,
s^2\, (\J / 2e) \, \coth (s \J /2e) \, {\rm e}^{-s } ~, \non \\
\G_{\rm VI}^{(e)} &= &
\frac{1}{(4\p)^4 }
\int {\rm d}^8 z \, 
\frac{W^2 {\bar W}^2}{(\f {\bar \f})^2}
\int \limits_{0}^{\infty}  {\rm d}s \,
\int \limits_{0}^{\infty}  {\rm d}t \, (s+t)^2\,
(\J / 2e) \, \coth (\J (s+t)/2e) \, {\rm e}^{-(s+t) } \non \\
&-& \frac{2}{(4\p)^4 }
\int {\rm d}^8 z \, 
\frac{W^2 {\bar W}^2}{(\f {\bar \f})^2}
\int \limits_{0}^{\infty}  {\rm d}s \,
(1+s+\hf s^2)\, (\J / 2e) \, \coth (s \J /2e) \, {\rm e}^{-s } ~, 
\label{VI} \\
\G_{\rm VII}^{(0)} &= &
-\frac{1}{(4\p)^4 }
\int {\rm d}^8 z \, 
\frac{W^2 {\bar W}^2}{(\f {\bar \f})^2}
\int \limits_{0}^{\infty}  {\rm d}s \,
\int \limits_{0}^{\infty}  {\rm d}t \, s^2\,
(\J / 2e) \, \coth (\J (s+t)/2e) \, {\rm e}^{-(s+t) } \non \\
&+& \frac{2}{(4\p)^4 }
\int {\rm d}^8 z \, 
\frac{W^2 {\bar W}^2}{(\f {\bar \f})^2}
\int \limits_{0}^{\infty}  {\rm d}s \,
(\J / 2e) \, \coth (s \J /2e) \, {\rm e}^{-s } ~, \non \\
\G_{\rm VII}^{(e)} &= &
 \frac{4}{(4\p)^4 }
\int {\rm d}^8 z \, 
\frac{W^2 {\bar W}^2}{(\f {\bar \f})^2}
\int \limits_{0}^{\infty}  {\rm d}s \,
(1+s +\hf s^2)\,
(\J / 2e) \, \coth (s \J /2e) \, {\rm e}^{-s } ~. 
\label{VII}
\eea

As concerns $\G_{\rm VIII}$, this is the only 
`eight' supergraph in the pure $\cN=2$ 
super Yang-Mills sector. 
Its direct evaluation leads to the following divergent 
contribution:
\bea
{ \G_{\rm VIII} \over N(N-1)} 
&=& -
 \frac{2}{(4\p)^4 }
\int {\rm d}^8 z \, 
\frac{W^2 {\bar W}^2}{(\f {\bar \f})^2}
\int \limits_{0}^{\infty}  {\rm d}s \, s\,
 (\J / 2e) ^2\,
\frac{  \cosh (s \J /e)}  { \sinh^2 (s \J /2e) } 
 \, {\rm e}^{-s } \non \\
&=& -
 \frac{2}{(4\p)^4 }
\int {\rm d}^8 z \, 
\frac{W^2 {\bar W}^2}{(\f {\bar \f})^2}
\int \limits_{0}^{\infty}  {\rm d}s \, s\,
\left\{ 
\frac{ (\J / 2e) ^2 }  { \sinh^2 (s \J /2e) } 
-{1 \over s^2} \right\}   \, {\rm e}^{-s } \non \\
&& - \frac{2}{(4\p)^4 }
\int {\rm d}^8 z \, 
\frac{W^2 {\bar W}^2}{(\f {\bar \f})^2}
\int \limits_{0}^{\infty} { {\rm d}s \over s} \,{\rm e}^{-s } 
-\frac{1}{(4\p)^4 }
\int {\rm d}^8 z \, 
\frac{W^2 {\bar W}^2}{(\f {\bar \f})^2} \, 
(\J/e)^2~. 
\eea
The divergence is now isolated, and is given by the second 
term. The first term can be integrated by parts using 
the identity (\ref{by-parts}).
This gives
\bea
{ \G_{\rm VIII} \over N(N-1)} 
&=& -
 \frac{2}{(4\p)^4 }
\int {\rm d}^8 z \, 
\frac{W^2 {\bar W}^2}{(\f {\bar \f})^2}
\int \limits_{0}^{\infty}  {\rm d}s \,
 (\J / 2e)\,  \coth (s \J /2e) \,{\rm e}^{-s } \non \\
&&+ \frac{2}{(4\p)^4 }
\int {\rm d}^8 z \, 
\frac{W^2 {\bar W}^2}{(\f {\bar \f})^2}
\int \limits_{0}^{\infty}  {\rm d}s \, s\,
\left\{  (\J / 2e)\, { \coth (s \J /2e) } 
-{1 \over s} \right\}  \,{\rm e}^{-s } \non \\
&&
-\frac{1}{(4\p)^4 }
\int {\rm d}^8 z \, 
\frac{W^2 {\bar W}^2}{(\f {\bar \f})^2} \, 
(\J/e)^2~. 
\label{VIII}
\eea
Here the divergence has been absorbed into the first term.

The quantum corrections (\ref{IV}) -- (\ref{VII}) and
(\ref{VIII})
produce a divergent $F^4$ contribution contained in 
\be 
\frac{1}{(4\p)^4 }
\int {\rm d}^8 z \, 
\frac{W^2 {\bar W}^2}{(\f {\bar \f})^2}
\int \limits_{0}^{\infty}  {\rm d}s \,
 (\J / 2e)\,  \coth (s \J /2e) \,{\rm e}^{-s }~.
\ee
But the total contribution of this functional form 
is multiplied by  
$$ 
1 -1 - 2 -2 + 2 +4 -2 = 0~,
$$
 and therefore no divergence is present.

It only remains to  convert the double proper-time 
integrals in (\ref{IV}) -- (\ref{VII}) into  single 
ones by implementing the change of 
variables (\ref{change-of-var}).
Then, the complete two-loop contribution 
(\ref{sym-two}) to the effective 
action from the pure $\cN=2$  super Yang-Mills sector, 
including the ghosts, is
\bea
{\G_{\rm SYM} \over N(N-1)}
&=& 
 \frac{3}{(4\p)^4 }
\int {\rm d}^8 z \, 
\frac{W^2 {\bar W}^2}{(\f {\bar \f})^2} 
-\frac{1}{(4\p)^4 }
\int {\rm d}^8 z \, 
\frac{W^2 {\bar W}^2}{(\f {\bar \f})^2} \, 
(\J/e)^2
\label{final-sym}
\\
&+&
\frac{1}{(4\p)^4 }
\int {\rm d}^8 z \, 
\frac{W^2 {\bar W}^2}{(\f {\bar \f})^2}
\int \limits_{0}^{\infty}  {\rm d}s \,
\Big( 4s - s^2 + s^3\Big)\,
\left\{
(\J / 2e)  \coth (s\J/2e) 
-{1\over s} \right\} \, {\rm e}^{-s } ~.\non
\eea

\sect{Conclusion}
Under the relaxed super self-duality condition, 
the two-loop effective action for  
the $\cN=4$ super Yang-Mills theory 
is the sum of the two contributions  
given in eqs. (\ref{final-hyper}) and 
(\ref{final-sym}). 
These results lead to the expression 
(\ref{O-two}) for
$ \O_{\rm two-loop}(\J^2, 0)$, 
upon restoring the explicit dependence on the coupling 
constant $g^2 = 2 g_{\rm YM}^2$.

It should be noted that we have not restricted ourselves
to the planar limit, in which $N\to \infty$. 
The expression (\ref{O-two})  for
$ \O_{\rm two-loop}(\J^2, 0)$ includes all subleading 
contributions, and such a result is of interest in itself.

As a consequence of  (\ref{O-two}), 
there is no two-loop $F^4$ quantum correction
for the $\cN=4$ $SU(N)$ theory, 
in accordance with the conjectures of \cite{DS,KM3}.
This is a nontrivial result. Our analysis above shows 
that non-vanishing $F^4$ 
contributions are generated by several two-loop 
supergraphs.  But their total contribution turns out to be zero.
In this respect, it is worth pointing out that 
a non-vanishing two-loop $F^4$ quantum 
correction does appear
in the case of $\cN=2$ $SU(N)$ 
SYM with $2N$ hypermultiplets in 
the fundamental \cite{KM3}.
The latter theory is a finite $\cN=2$
supersymmetric field theory, 
but it possesses no supergravity 
dual in the framework of the AdS/CFT 
correspondence.

It follows from the analysis of two-loop 
supergraphs that, at intermediate stages,  
there appear contributions
to $ \O_{\rm two-loop}(\J^2, 0)$ of the form
\be
\int \limits_{0}^{\infty}  {\rm d}s \,
\sum\limits_{n=0}^{3} c_n \, s^n
\left\{
(\J / 2e)  \coth (s\J/2e) 
-{1\over s} \right\} \, {\rm e}^{-s } ~, 
\ee
with $c_n$ numerical coefficients.
In the final expression for $ \O_{\rm two-loop}(\J^2, 0)$, 
eq.  (\ref{O-two}), it is only the coefficient $c_2$ 
which does not vanish. It is easy to understand why 
$c_0 = 0$ -- this is equivalent to the cancellation 
of divergences which are present  in some (and only in) 
$F^4$ quantum corrections.
We do not understand, however, 
why $c_1 $ and  $c_3 $ vanish.
In the case of $\cN=2$ super QED, 
one has $c_1 = 0$ and $c_2 = -3 \,c_3 \neq 0$ 
at two loops \cite{KM4}.

An unexpected outcome of our analysis
is that no $F^6$ quantum correction occurs at two loops, 
and this is in contradiction with the $\cN=2$ harmonic superspace
calculation of \cite{BPT}. {\it If} our calculation is 
correct\footnote{It is worth pointing out 
that  our work   is the first calculation of such complexity 
in supersymmetric gauge theories.}
(all the technical steps of the calculation
have been checked several times), 
then there should be a reason why 
no $F^6$ quantum corrections appear both at one 
and two loops.\footnote{One hint that a 
non-vanishing $F^6$ term may be expected 
is the agreement \cite{BBPT} between the 
$v^6 / |X|^{14}$ term in the interaction 
potential between D0-branes in the supergravity 
description and the corresponding two-loop term in 
the effective action of the Matrix model.}  
Although we presently  have no solid
explanations, it is worth speculating about 
possible reasons.

The $F^6$ term turns out to be rather special 
from the point of view of extended supersymmetry.
In $\cN=3$ harmonic superspace 
(see \cite{GIOS} for a review), 
one can construct super-extensions
of component structures $F^{4k}$, where $k=1,2, \dots$, 
in terms of  the $\cN=3$ superfield strengths, 
but no manifestly supersymmetric extension 
of the $F^6$ term  exists 
\cite{IZ}.
As shown by Ivanov and Zupnik \cite{IZ}, a non-vanishing 
$F^6$ term is then  generated at the component level
(provided the $\cN=3$ supersymmetric $F^4$ term is present) 
 only upon elimination of some bispinor auxiliary fields
belonging to the off-shell $\cN=3$ vector multiplet.
The $\cN=3$ SYM theory is known to be classically equivalent 
to  the $\cN=4$ SYM theory \cite{GIOS}, but only three
supersymmetries are manifestly realized 
in the $\cN=3$ harmonic superspace formulation \cite{GIOS}.
In the case of $\cN=4$ SYM, when one tries to keep control 
of  an additional supersymmetry, it is clear that the constraints 
imposed by $\cN=3$ supersymmetry cannot be relaxed, but 
new constraints may appear.

Independent of the $\cN=3$ harmonic superspace 
story, it would be interesting to understand whether 
there exists an $\cN=4$ supersymmetric completion 
of the $F^6$ term, both in the $\cN=1$ and $\cN=2$ 
superspace settings.\footnote{The low-energy effective action 
for the $\cN=4$ SYM theory should possess some kind
of  (deformed) $\cN=4$ on-shell supersymmetry
provided one retains all 
($\cN=0$, or $\cN=1$, or $\cN=2$)
components of the $\cN=4$ 
vector multiplet.}    
Such an analysis seems to be easier
in $\cN=2$ superspace,
where the $F^6$ term is
\cite{BKT}
\be
\int{\rm d}^4 x \, {\rm d}^8 \q \, 
{1 \over {\bar W}^2 } \, 
\ln W \, D^4 \ln W~,
\ee
compare with (\ref{n=1-F^6}).
To obtain an $\cN=4$ vector multiplet, we should add 
to $W$ and $\bar W$ 
a single hypermultiplet described by some constrained 
$\cN=2$  superfields $Q^i$ and ${\bar Q}_i$, where $i=1,2$
(or, in the $\cN=2$ harmonic superspace approach 
\cite{GIOS}, by unconstrained 
analytic superfields $Q^+$ and $\tilde{Q}^+$). 
Now, the idea is to look for a completion of
this functional by hypermultiplet
dependent contributions
such that the resultant functional is invariant 
under two additional supersymmetries
(a similar problem has recently been solved for 
the $F^4$ term in \cite{BI}).
If this proves to be impossible to realize, 
then indeed the absence of  $F^6$ quantum corrections
should be natural.

The problem of $\cN=4$ on-shell completion of the 
$\cN=2$ superconformal  $F^6$, $F^8$ and higher order 
terms \cite{BKT} has recently been analyzed by 
Banin and Pletnev \cite{BP}. Regarding the $F^6$ term, 
their conclusion is that its  $\cN=4$ on-shell completion  
 is feasible provided (i) a non-vanishing $F^4$ term is present; 
(ii) $\cN=4$ supersymmetry is deformed (as compared 
with the supersymmetry transformations of the classical action)
in a special manner. So, the fate of the $F^6$ term depends 
on the explicit structure of the quantum deformation of two hidden supersymmetries. If the quantum theory supports the deformation 
advocated in \cite{BP},  then there must be a non-vanishing 
$F^6$ term in the effective action; otherwise it must vanish. 
But the issue of quantum supersymmetry deformation
in the $\cN=4$ super Yang-Mills theory
is still open.

In our opinion, 
the absence of  two-loop $F^6$ contributions to the 
$\cN=4$ SYM  effective action 
does not  mean that 
the conjectured correspondence 
\cite{Maldacena,CT,Tseytlin,BPT}
between the D3-brane action in $AdS_5 \times S^5$
and the low-energy action for $\cN=4$ $SU(N)$ 
SYM (on its Coulomb branch)  is in doubt. 
Rather, it  is an indication that this correspondence  
is more subtle than previously  thought. 
The vanishing of the two-loop $F^6$ contribution 
involves the cancellation of two terms, in eq. (\ref{O-two}),
one of which generates the Born-Infeld $F^6$ term. 
These two contributions have very 
different origins at the level  of supergraphs, 
and the one which generates the Born-Infeld $F^6$ 
term  (the first term on the right of (\ref{O-two}))
is not accompanied  by higher powers of $F$. 
This would be consistent with  
an idea that the superconformal 
Born-Infeld action (\ref{BI}) 
should correspond {\it only} to a sub-sector of
the $\cN=4$ SYM effective action
in the large $N$, fixed $N g_{\rm YM}^2 \gg 1$ 
approximation.

In the large $N$ limit, the expression 
(\ref{O-two})  takes the form
\bea
 \O_{\rm two-loop}(\J^2, 0) 
= - {(2g_{\rm YM}^2 N)^2 \over (4 \p)^4 }  
\left( \J^2  
-2 \int_{0}^{\infty} {\rm d} s\, s^2 \Big\{ 
{ \J \over 2  } \coth {s  \J \over 2} - {1 \over s} 
\Big\}  \, {\rm  e}^{-s} \right)~+~O(N)~.
\non 
\eea
It generates quantum corrections of the form 
$F^{8+2k}$, where $k=0,1,\dots $, proportional to 
$\l^2$, with  $\l =g_{\rm YM}^2 N $.
This result is in fact in accord with the conjecture 
of  \cite{BPT}.  As applied to the $F^8$ 
terms,\footnote{One of us (SMK) was informed by 
I. Buchbinder and A. Tseytlin that, 
together with A. Petrov and O. Solomina,
they had computed two-loop $F^8$ corrections 
for the $\cN=4$ $SU(N)$ SYM theory. 
Their $F^8$ results have the same qualitative
structure as ours, but there seem to be quantitative differences.}  
the conjecture (i) allows  sub-leading
one- and two -loop contributions proportional 
to $\l$ and $\l^2$;  (ii) requires the dominant contribution, 
proportional to $\l^3$, to appear at three loops.

The $F^6$ puzzle raised in our paper seems to be similar in nature 
to that uncovered by Stieberger and Taylor \cite{ST} who 
computed the two-loop perturbative $F^6$ interactions 
in $SO(32)$ heterotic supersting theory and compared them,
guided by heterotic -- type I string duality \cite{Witten3,PW},  
with those predicted by the expansion of (any form of)  
non-Abelian generalization of the Born-Infeld action 
(see \cite{Tseytlin} for a review). Quite unexpectedly,
even when restricted to the Cartan subalgebra of $SO(32)$, 
their two-loop action does not agree with the weak 
field expansion of the Abelian Born-Infeld action. 
As argued in \cite{ST}, what makes $F^6$ truly different 
from $F^4$ is that the corresponding amplitudes are not 
``BPS-saturated.'' This indicates that the $F^6$ terms 
are sensitive to the full spectrum of superstring theory,
and therefore the comparison of dual descriptions may be  
rather nontrivial.

We believe that this paper sheds some light on the structure 
of perturbative expansions in the $\cN=4$ super Yang-Mills
theory on its Coulomb branch, the issue recently raised 
by Witten \cite{Witten2}.  What is more certain is that our  paper
raises more questions than  answers.

\vskip.5cm

\noindent
{\bf Acknowledgements:}\\
We are grateful to Ioseph Buchbinder 
and especially to  Arkady Tseytlin for reading the 
manuscript and for helpful critical comments. 
We are also grateful to Nikolay Pletnev for bringing 
the papers \cite{BP} to our attention.
We also thank the referee of this paper for bringing 
important references \cite{ST} to our attention.
This work is supported in part by the Australian Research
Council and  UWA research grants.

\begin{appendix}

\sect{Parallel displacement propagator}
In this appendix we describe, following \cite{KM}, 
the salient properties of the $\cN=1$ 
parallel displacement propagator
$I(z,z')$. This object is uniquely specified by 
the following requirements:\\
(i) the gauge transformation law
\be
 I (z, z') ~\to ~
{\rm e}^{{\rm i} \t(z)} \,  I (z, z') \,
{\rm e}^{-{\rm i} \t(z')} ~
\label{super-PDO1}
\ee
with respect to  an  arbitrary gauge ($\t$-frame) transformation 
of  the covariant derivatives
\be
\cD_A ~\to ~ {\rm e}^{{\rm i} \t(z)} \, \cD_A\,
{\rm e}^{-{\rm i} \t(z)}~, \qquad 
\t^\dagger = \t ~, 
\label{tau}
\ee
with the gauge parameter $\t(z)$ being arbitrary modulo 
the reality condition imposed;\\
(ii) the equation  
\be
\z^A \cD_A \, I(z,z') 
= \z^A \Big( D_A +{\rm i} \, \G_A(z) \Big) I(z,z') =0~;
\label{super-PDO2}
\ee
(iii) the boundary condition 
\be 
I(z,z) ={\bf 1}~.
\label{super-PDO3}
\ee
These imply the important relation
\be
I(z,z') \, I(z', z) = {\bf 1}~,
\label{collapse}
\ee
as well as 
\be
\z^A \cD'_A \, I(z,z') 
= \z^A  \Big( D'_A \,I(z,z') 
 - {\rm i} \, I(z,z') \, \G_A(z') \Big) =0~.
\ee
Under Hermitian conjugation, the parallel 
displacement propagator transforms as 
\be
\Big( I(z,z')\Big)^\dagger = I(z',z)~.
\ee

${}$For a covariantly constant vector multiplet, 
\be
\cD_a \, W_\b =0~, 
\label{constant SYM}
\ee 
the covariant differentiation of  $\cD_A \,I(z,z')$ gives \cite{KM}
\bea
\cD_{\b \bd} I(z,z') &=& I(z,z') \,
\Big( -  \frac{{\rm i}}{4} \r^{\ad \a} \cF_{\a \ad, \b \bd}(z') 
-{\rm i} \, \z_{\b} \bar{\cW}_{\bd}(z') 
+ {\rm i} \, \bar{\z}_{\bd} \cW_{\b}(z')
\nonumber \\ 
&& \phantom{I(z,z') \,\Big( }
+\frac{2{\rm i}}{3} \, 
\bar{\z}_{\bd}
\z^{\a}
\cD_{\a}\cW_{\b}(z')
+ \frac{2{\rm i}}{3} \, \z_{\b} \bar{\z}^{\ad}
\bar{\cD}_{\ad} \bar{W}_{\bd}(z') \Big) \nonumber \\
&=& \Big( - \frac{{\rm i}}{4} \r^{\ad \a} \cF_{\a \ad, \b \bd}(z) 
-{\rm i} \, \z_{\b} \bar{\cW}_{\bd}(z) 
+ {\rm i} \, \bar{\z}_{\bd} 
\cW_{\b}(z)
\nonumber \\ 
&  & 
\phantom{\Big(}
-\frac{{\rm i}}{3} \, \bar{\z}_{\bd} \z^{\a} \cD_{\a}\cW_{\b}(z) 
- \frac{{\rm i}}{3} \, \z_{\b} \bar{\z}^{\ad}
\bar{\cD}_{\ad} \bar{\cW}_{\bd}(z) \Big) \, I(z,z')~;
\label{super-PTO-der2} \\
\cD_{\b} I(z,z') &=& I(z,z') \,
\Big( \frac{1}{12} \, \bar{\z}^{\bd} 
\r^{\a \ad}  \cF_{\a \ad, \b \bd}(z') 
- {\rm i} \, \r_{\b \bd} \Big\{
\frac{1}{2} 
\bar{\cW}^{\bd}(z') 
-\frac{1}{3}  
\bar{\z}^{\ad} \bar{\cD}_{\ad}
\bar{\cW}^{\bd}(z')  \Big\}
\nonumber \\ 
&  & 
+\frac13 \, \z_{\b} \bar{\z}_{\bd} \bar{W}^{\bd}(z')  
 +\frac13
\bar{\z}^2 \Big\{ 
\cW_{\b}(z') + \hf \z^{\a} \cD_{\a}
\cW_{\b}(z')
-\frac{1}{4}  \z_{\b} \cD^{\a} \cW_{\a}(z') \Big\}
\Big)
\nonumber \\
&=& \Big( \frac{1}{12} \, \bar{\z}^{\bd} 
\r^{\a \ad} \, \cF_{\a \ad , \b \bd}(z)
- \frac{{\rm i}}{2} \, \r_{\b \bd} 
\Big\{ 
\bar{\cW}^{\bd}(z) 
+\frac{1}{3}  \bar{\z}^{\ad} \bar{\cD}_{\ad}
\bar{\cW}^{\bd}(z) \Big\} 
+ \frac13 \, \z_{\b} \bar{\z}_{\bd} \bar{\cW}^{\bd}(z)  \nonumber \\ 
& & \phantom{ \Big( } +\frac13
\bar{\z}^2 \Big\{
\cW_{\b}(z) - \hf  \z^{\a} \cD_{\a} \cW_{\b}(z)
+\frac{1}{4}  \z_{\b} {\cD}^{\a} {\cW}_{\a}(z) \Big\}
\Big)\, I(z,z')~;
\label{super-PTO-der3} \\
{\bar \cD}_{\bd} I(z,z') &=& I(z,z') \,\Big( 
-\frac{1}{12} \, \z^{\b } 
\r^{\a \ad} \, F_{\a \ad , \b \bd}(z')
- {\rm i} \,\r_{\b \bd} \Big\{ 
\frac{1}{2}  \cW^{\b}(z') 
+\frac{1}{3}  \z^{\a} \cD_{\a} \cW^{\b}(z')  \Big\}
\nonumber \\ &  & 
- \frac13 \, \bar{\z}_{\bd} \z^{\b} \cW_{\b}(z')  
-\frac13 \z^2 \Big\{ 
\bar{\cW}_{\bd}(z')  -\hf \bar{\z}^{\ad} \bar{\cD}_{\ad}
\bar{\cW}_{\bd}(z')
+\frac{1}{4}  \bar{\z}_{\bd} {\bar \cD}^{\ad} 
{\bar \cW}_{\ad}(z') \Big\} 
\Big)
\nonumber \\
&=& \Big( - \frac{1}{12} \, \z^{\b} 
\r^{\a \ad}  \cF_{\a \ad , \b \bd}(z)
- \frac{{\rm i}}{2} \, \r_{\b \bd} \Big\{ \cW^{\b}(z) 
- \frac{1}{3}  \z^{\a} \cD_{\a} \cW^{\b}(z)  \Big\} 
- \frac13 \, \bar{\z}_{\bd} \z^{\b} \cW_{\b}(z)  \nonumber \\ 
& & - \frac13 \z^2 \Big\{ 
\bar{\cW}_{\bd}(z) +\hf  \bar{\z}^{\ad} \bar{\cD}_{\ad}
\bar{\cW}_{\bd}(z)
-\frac{1}{4}  \bar{\z}_{\bd} {\bar \cD}^{\ad} 
{\bar \cW}_{\ad}(z) \Big\}
\Big) \, I(z,z')~.
\label{super-PTO-der4} 
\eea

\sect{\mbox{$U(1)$}
heat kernel 
 in a self-dual background}
In this appendix, we describe 
simplifications in the structure 
of the $U(1)$ heat kernel
\bea
\bK^{(e)}(z,z'|s) &=& -\frac{\rm i}{(4 \pi s)^2} \, 
\sqrt{  \det
\left( \frac{s \,eF}{\sinh  (s\, eF )}\right) } \; 
\d^2 (\z^{(e)} (s) ) 
\d^2( \bar{\z}^{(e)} (s)) \non \\
& \times &
 {\rm U}(s) \,
{\rm e}^{ \frac{{\rm i}}{4} 
\r \, eF \coth ( s \, eF) \, \r } \, I(z,z') ~.
\label{U(1)kernel}
\eea
with
\bea 
\z^{(e)\,\a}(s) 
&=& {\rm U}(s) \, \z^\a\, {\rm U}(-s)~, 
\qquad 
{\bar \z}^{(e)\,\ad}(s) 
= {\rm U}(s) \, {\bar \z}^\ad\, {\rm U}(-s)~, \non \\
{\rm U}(s) &=&  \exp \Big\{- {\rm i} s \,e( W^{\a} \cD_{\a} 
+  \bar{W}^{\ad} {\bar \cD}_{\ad})\Big\}~,
\eea
which occur when the background vector 
multiplet satisfies the relaxed self-duality condition 
\be
W_\a \neq 0~, \quad 
D_\a W_\b = 0~, \qquad 
{\bar D}_{(\ad} {\bar W}_{\bd)} \neq 0~.
\ee
With the notation
\be
{\bar N}_\ad{}^\bd = {\bar D}_\ad {\bar W}^\bd~, 
\qquad 
{\bar B}^2 = \hf \,{\rm  tr}  {\bar N}^2 = {1 \over 4}
{\bar D}^2 {\bar W}^2 ~,
\ee
one obtains
\be
\sqrt{  \det
\left( \frac{s \,eF}{\sinh  (s\, eF )}\right) } 
= \left( \frac{ s\,e{\bar B}/2}
{\sin ( s\,e{\bar B}/2)} \right)^2
\equiv \L(s\,e{\bar B}/2)
~. 
\label{lambda}
\ee
Since the field strength is self-dual, 
the matrix $eF \coth ( s \, eF)$, which  
appears in the exponential in (\ref{U(1)kernel}),
becomes
a multiple of the unit matrix, 
\be
eF \coth ( s \, eF)
= \frac{e \bar B }{2} \,
\cot (s \,e {\bar B} /2 ) \, {\bf 1}_4~.
\ee

One also obtains
\bea
\z^{(e) \,\a}(s) 
= \z^{\a} - {\rm i}s\, eW^\a~,
\qquad
{\bar \z}^{(e)\, \ad}(s) 
= {\bar \z}^{\ad} - \Big(  {\bar W} \,
\frac{  {\rm e}^{-{\rm i}s e{\bar N}} -1} {\bar N}\Big)^{\ad}~.
\label{shifted}
\eea

\sect{Proof of some identities I}

In this appendix, we illustrate the types of manipulations 
which can be
performed on the functional expressions for supergraphs 
by proving  the identities
(\ref{A2}) and (\ref{A3}). These manipulations have also been used 
extensively  in section 4 to bring the
expressions for individual supergraphs into a relatively standard form.

Let us denote by $I$ the  right hand side of (\ref{A2}),
$$ I = \int {\rm d}^8 z   \int {\rm d}^8 z' \,
G^{\m \n}(z,z') \,{\rm tr}_{\rm Ad}  \Big(
T_\m  \, \cD^\a G(z,z')  \, T_\n \, \cW'_\a \,
\cD'^2 {\bar \cD}'^2 G(z',z) \Big) ~.$$
One can use the identitity (\ref{A1}) to
replace $\cD^\a G(z,z')$ by $-\cD'^\a G(z,z').$ 
Integrating  by parts
with respect to the derivative $\cD'^\a,$ one arrives at
\bea I &=& \int {\rm d}^8 z   \int {\rm d}^8 z' \,
(\cD'^\a G^{\m \n}(z,z') ) \,{\rm tr}_{\rm Ad}  \Big(
T_\m  \,  G(z,z')  \, T_\n \, \cW'_\a \,
\cD'^2 {\bar \cD}'^2 G(z',z)  
 \Big)
\nonumber \\
&=&  \int {\rm d}^8 z   \int {\rm d}^8 z' \,
( \cD'^\a G^{\m \n}(z,z'))
 ( T_{\m})^{\rho}{}_{\sigma} \, G^{\sigma}{}_{\gamma}
(z,z') \, \,  (T_{\n})^{\gamma}{}_{\delta} \, (\cW'_\a \,
\cD'^2 {\bar \cD}'^2 G(z',z) )^{\delta}{}_{\rho},
\eea
where the matrices  $(T_\m )^\l{}_\n$
in the adjoint representation are  defined by
\be
[\cT_\m ,  \cT_\n ] = \cT_\l \,
(T_\m )^\l{}_\n ~,
\ee
with $\cT_m$ the generators of 
$SU(N)$ in any representation.
This can be re-arranged as
\bea
I &=& - \int {\rm d}^8 z   \int {\rm d}^8 z' \,
G^{\sigma \g}
 (T_{\sigma})^{\rho}{}_{\m} \, \cD'^\a G^{\m}{}_{ \n}(z,z') \,  
(T_{\gamma})^{\n}{}_{\delta} \, (\cW'_\a \,
\cD'^2 {\bar \cD}'^2 G(z',z) )^{\delta}{}_{\rho} \nonumber \\
& = &  \int {\rm d}^8 z   \int {\rm d}^8 z' \,
G^{\sigma \gamma}(z,z') \,{\rm tr}_{\rm Ad}  \Big(
T_{\sigma}  \, \cD'^\a  G(z,z')  \, T_{\gamma} \, \cW'_\a \,
\cD'^2 {\bar \cD}'^2 G(z',z) \Big) ~. \label{A4}
\eea
Using (\ref{A1}) again, the right hand side of (\ref{A4}) is 
equivalent to $-I,$
proving that $I$ vanishes.

To prove (\ref{A3}), one begins with the identity
\bea
0 &=& \int {\rm d}^8 z   \int {\rm d}^8 z' \, \cD'^\a
\Big( G^{\m \n}(z,z') \,{\rm tr}_{\rm Ad} ( \, T_\m \, \cD^\b \,
G(z,z') \, \cW'_\b \, T_\n \, \cD'_\a {\bar \cD}'^2 G(z', z) ) \Big)
\nonumber \\
&=& \int {\rm d}^8 z   \int {\rm d}^8 z' \, \Big\{
- (\cD^\a G^{\m \n} (z,z')) \,{\rm tr}_{\rm Ad}  \Big( T_\m \, \cD^\b \,
G(z,z') \, \cW'_\b \, T_\n \, \cD'_\a {\bar \cD}'^2 G(z', z) \Big)
\nonumber \\
&&\qquad \quad + 
 \frac12 \, G^{\m \n} (z,z') \, {\rm tr}_{\rm Ad}  \Big( T_\m \,
\cD^2 \,
G(z,z') \, \cW'_\a \, T_\n \, \cD'_\a {\bar \cD}'^2 G(z', z) \Big)
\nonumber  \\
&& \qquad \quad +~
 G^{\m \n}(z,z') \,{\rm tr}_{\rm Ad}  \Big( T_\m \, \cD^\b \,
G(z,z') \, \cW'_\b \, T_\n \, \cD'^2 {\bar \cD}'^2 G(z', z) \Big) \Big\} ~ .
\label{A5}
\eea
The first term on the right hand side of (\ref{A5}) is
$$
- \int {\rm d}^8 z   \int {\rm d}^8 z' \,
  (\cD^\a G^{\m \n} (z,z')) \, (T_\m)^\o{}_\r \, \cD^\b \,
G(z,z')^\r{}_\e \, (\cW'_\b )^\e{}_\s \, (T_\n)^\s{}_\t \,
\cD'_\a {\bar \cD}'^2 G(z', z)^\t{}_\o ~ .
$$
Using the commutation relations 
and the
symmetry properties of the generators in the adjoint representation,
this can be re-expressed in the form
$$
- \int {\rm d}^8 z   \int {\rm d}^8 z' \, (\cD'^\b G^{\m \n} (z,z')) 
\,{\rm tr}_{\rm Ad}  \Big( T_\m \, \cD^\a \,
G(z,z') \, [T_\n ,\cW'_\b ]\, \cD'_\a {\bar \cD}'^2 G(z', z) \Big) ~.
$$
Integrating by parts with respect to the derivative $\cD'^\b$ yields
the expression
\bea
& {} &\int {\rm d}^8 z   \int {\rm d}^8 z' \,
  G^{\m \n} (z,z') \, {\rm tr}_{\rm Ad}  \Big(- \frac12 \, T_\m \,
  \cD^2 \,
G(z,z') \, [T_\n ,\cW'^\a ]\, \cD'_\a {\bar \cD}'^2 G(z', z)\non  \\
&+& T_\m \, \cD^\a \,
G(z,z') \, [T_\n ,\cW'_\a ]\, \cD'^2 {\bar \cD}'^2 G(z', z) \Big).
\eea
Substituting  this expression for the first term in (\ref{A5}) back in, 
and using
the identity (\ref{A2}), yields the required result (\ref{A3}).

Other identities given in section 4 can be proved by 
similar means.

\sect{Proof of some identities II}

In this appendix, details of the group-theoretical 
manipulations leading to the expressions 
(\ref{fish-structure-0}) and (\ref{eight-structure-0}) 
are presented. They make use of the following identities for 
the generators of $SU(N)$ in the adjoint representation: 
\bea
\sum_{I}\, \sum_{\iu ,\ju} \, (H_I)^{0 \iu}{}_{0 \ju} \, (H_I)^{0 
\ju}{}_{0 \iu} &=& 2(N-1)~,
\label{id1} \\
\sum_{\iu\neq  \ju} \, \sum_{\ku, \lu} \, (E_{\iu \ju})^{0 \ku}{}_{0 \lu} \, 
(E_{\ju \iu})^{0 \lu}{}_{0 \ku} &=& (N-1)(N-2)~,
\label{id2} \\
\sum_{\iu, \ju} \, \sum_{I} \, (E_{0 \iu})^{0 \ju}{}_{I} \, 
(E_{\iu 0})^{I}{}_{0 \ju} &=& 2 (N-1) ~, 
\label{id3} \\
\sum_{\iu, \ju} \, \sum_{\ku, \lu} \, (E_{0 \iu})^{\ku \lu}{}_{\ju 0} \, 
(E_{\iu 0})^{\ju 0}{}_{\ku \lu} &=& (N-1) (N-2)~,
\label{id4} \\
\sum_{I} \, \sum_{\iu, \ju} \, \sum_{\ku, \lu} \, (H_I)^{\iu 
\ju}{}_{\ku \lu} \, (H_I)^{\ku 
\lu}{}_{\iu \ju} &=& 2 (N-1) (N-2) ~,
 \label{id6}\\
\sum_{\iu \neq \ju} \, \sum_{\ku, \lu} \, \sum_{\pu, \qu} (E_{\iu 
 \ju})^{\ku 
\lu}{}_{\pu \qu} \, (E_{\ju \iu})^{\pu 
\qu}{}_{\ku \lu} &=& 2 (N-1) (N-2) (N-3)
~, 
\label{id7}\\
\sum_{\iu, \ju} \, (E_{0 \iu} \, E_{\iu 0})^{0 \ju }{}_{0 \ju 
} &=& N (N-1)~,
\label{id5} \\
\sum_{\iu, \ju}  \, (E_{0 \iu} \, E_{\iu 0})^{\ju 0}{}_{\ju 
0} &=& 0~,
\label{id8} \\
\sum_{I} \,  \sum_{\iu} \, (H_I \, H_I)^{0 \iu }{}_{0 \iu } 
&=& 2 (N-1)~,
\label{id9} \\
\sum_{I} \,  \sum_{\iu, \ju} \, (H_I \, H_I)^{\iu \ju }{}_{\iu \ju } 
&=& 2  (N-1) (N-2)~,
\label{id10}  \\
\sum_{\iu \neq  \ju}\,  \sum_{\ku} \, (E_{\iu \ju} \, 
E_{\ju \iu})^{\ku 0}{}_{\ku 0} 
= \sum_{\iu \neq \ju} \sum_{\ku} (E_{\iu \ju} \, E_{\ju \iu})^{0 
\ku}{}_{0 \ku } &=& (N-1) (N-2)~.
\label{id11}
\eea
The above idenitities are easily proved by using the definition 
$$ [T_{\mu}, T_{\nu}] = T_{\rho} \, (T_{\mu})^{\rho}{}_{\nu}$$ of 
the  adjoint representation  and evaluating the commutators 
in the fundamental  representation,  
which is detailed in section 5.

Using (\ref{G}) and (\ref{I+II-ad12}), the expression 
(\ref{Gfish}) for the generic `fish' supergraph 
can be decomposed as
\bea
\G_\ominus & = & \int {\rm d}^8 z   \int {\rm d}^8 z' \,  \Big( \,
\bG^{(e)}(z,z') \, \sum_{\iu} \, {\rm tr}_{\rm Ad} \,    
( E_{0 \iu} \,  \hat{\cG}(z,z') \, E_{\iu 0} \,
\check{\cG}(z',z) ) \nonumber \\
& & \quad + \, \bG^{(- e)}(z,z') \, \sum_{\iu} \,{\rm tr}_{\rm Ad} \,    
( E_{\iu 0} \,  \hat{\cG}(z,z') \, E_{0 \iu} \,
\check{\cG}(z',z) ) \nonumber \\
& & \quad + \, \bG^{(0)}(z,z') \, \sum_{I} \,{\rm tr}_{\rm Ad} \,    
( H_I \,  \hat{\cG}(z,z') \, H_I \,
\check{\cG}(z',z) ) \nonumber \\
& & \quad + \, \bG^{(0)}(z,z') \, \sum_{\iu \neq \ju} \,{\rm tr}_{\rm Ad} \,    
( E_{\iu \ju} \,  \hat{\cG}(z,z') \, E_{\ju \iu} \,
\check{\cG}(z',z) ) \, \Big)\nonumber \\
& \equiv  & \G_\ominus^{(1)} + \G_\ominus^{(2)} 
+ \G_\ominus^{(3)} +  \G_\ominus^{(4)}~.
\label{fish1}
\eea

The calculation of $\G_\ominus^{(1)}$ will be outlined. 
The remaining  terms in $\G_\ominus$ follow in a similar manner. 
Due to the presence 
of $\bG^{(e)}$ in $\G_\ominus^{(1)}$,  
the considerations of
$U(1)$ charge conservation at each vertex  
dictate that either $\hat{\cG}$ has charge
 $-e$ and $\check{\cG}$ has 
charge 0, or $\hat{\cG}$ has charge 0 and $\check{\cG}$ has 
charge $e$. As a result,
\bea
\G_\ominus^{(1)} &=& \int {\rm d}^8 z   \int {\rm d}^8 z' \,  
\bG^{(e)}(z,z') \, \Big( \sum_{I} \, \sum_{\iu, \ju} \, (E_{0 
\iu})^I{}_{\ju 0} \,  \hat{\bG}^{(-e)}(z,z') \, (E_{ \iu 0})^{\ju 
0}{}_I \, \check{\bG}^{(0)}(z',z) \nonumber \\
& & \quad + \, \sum_{\iu, \ju, \ku, \lu} \, (E_{0 
\iu})^{\ku \lu}{}_{\ju 0} \,  \hat{\bG}^{(-e)}(z,z') \, (E_{ \iu 0})^{\ju 
0}{}_{\ku \lu} \, \check{\bG}^{(0)}(z',z) \nonumber \\
& & \quad + \, \sum_{I} \,\sum_{\iu, \ju} \, (E_{0 
\iu})^{0 \ju}{}_{I} \,  \hat{\bG}^{(0)}(z,z') \, (E_{ \iu 
0})^{I}{}_{0 \ju} \, \check{\bG}^{(e)}(z',z) \nonumber \\
& & \quad + \, \sum_{\iu, \ju, \ku, \lu} \, (E_{0 
\iu})^{0 \ju}{}_{\ku \lu} \,  \hat{\bG}^{(0)}(z,z') \, (E_{ \iu 
0})^{\ku \lu}{}_{0 \ju} \, \check{\bG}^{(e)}(z',z) \, \Big)~.
\eea
Making use of the identities (\ref{id3}) and  (\ref{id4}), this yields
\bea
\G_\ominus^{(1)} &=& N (N-1) \, \int {\rm d}^8 z   \int {\rm d}^8 z' \,  
\bG^{(e)}(z,z') \non \\
&& \quad \times
\Big( \,\hat{\bG}^{(-e)}(z,z') \, 
\check{\bG}^{(0)}(z',z) 
+ \, \hat{\bG}^{(0)}(z,z') \, 
\check{\bG}^{(e)}(z',z) \, \Big)~.
\eea

In a similar manner, $\G_\ominus^{(2)},$ $\G_\ominus^{(3)} $ and 
$\G_\ominus^{(4)}$ can be computed using the above identities:
\bea
\G_\ominus^{(2)} &=& N (N-1) \, \int {\rm d}^8 z   \int {\rm d}^8 z' \,  
\bG^{(-e)}(z,z') \non \\
& &  \times   \Big( \,\hat{\bG}^{(e)}(z,z') \, 
\check{\bG}^{(0)}(z',z) + \hat{\bG}^{(0)}(z,z') \, 
\check{\bG}^{(-e)}(z',z) \, \Big) ~, \nonumber \\
\G_\ominus^{(3)} &=& 2  (N-1) \, \int {\rm d}^8 z   \int {\rm d}^8 z' \,  
\bG^{(0)}(z,z') \, \Big(  \, \hat{\bG}^{(e)}(z,z') \, 
\check{\bG}^{(e)}(z',z)  \nonumber \\ 
& & \quad +  \, \hat{\bG}^{(-e)}(z,z') \, 
\check{\bG}^{(-e)}(z',z) + (N-2) \, \hat{\bG}^{(0)}(z,z') \, 
\check{\bG}^{(0)}(z',z) \, \Big) \nonumber \\
\G_\ominus^{(4)} &=&(N-1) (N-2) \, \int {\rm d}^8 z   \int {\rm d}^8 z' \,  
\bG^{(0)}(z,z') \, \Big( \,\hat{\bG}^{(e)}(z,z') \, 
\check{\bG}^{(e)}(z',z) \nonumber \\
& & \quad + \, \hat{\bG}^{(-e)}(z,z') \, 
\check{\bG}^{(-e)}(z',z)  + 2 (N-1) \, \hat{\bG}^{(0)}(z,z') \, 
\check{\bG}^{(0)}(z',z)\, \Big) 
\eea
Together with $\G_\ominus^{(1)},$ these yield 
the final result (\ref{fish-structure-0}) for $\G_\ominus.$

We now  turn  to the derivation of the result 
(\ref{eight-structure-0}) for the generic  `eight' supergraph. Using 
(\ref{hatG}), the expression (\ref{Geight}) for these supergraphs
 can be decomposed as
\bea
\G_\infty &= & \int {\rm d}^8 z  
\lim_{z' \to  z} \, \Big(
\hat{\bG}^{(e)}(z,z') \,\sum_{\iu} \, {\rm tr}_{\rm Ad}
( E_{0 \iu} \, E_{\iu 0} \, \check{\cG}(z,z') ) \nonumber \\
& & \quad + \, \hat{\bG}^{(-e)}(z,z') \,\sum_{\iu} \, {\rm tr}_{\rm Ad}
(  E_{\iu 0} \, E_{0 \iu} \, \check{\cG}(z,z') ) \nonumber \\
& & \quad + \, \hat{\bG}^{(0)}(z,z') \,\sum_{I} \, {\rm tr}_{\rm Ad}
(  H_I \, H_I \, \check{\cG}(z,z') ) \nonumber \\
& & \quad + \, \hat{\bG}^{(0)}(z,z') \,
\sum_{\iu \neq \ju} \, {\rm tr}_{\rm Ad}
(  E_{\iu \ju} \, E_{\ju \iu} \, \check{\cG}(z,z') ) \, \Big) \nonumber \\
& \equiv & \G_\infty^{(1)} + \G_\infty^{(2)} + \G_\infty^{(3)} + 
\G_\infty^{(4)}~.
\eea

Again, we detail only the calculation of $\G_\infty^{(1)}$.  The 
remaining terms in $\G_\infty$ follow by similar manipulations. 
Expanding out the trace,
\bea
\G_\infty^{(1)} &=& \int {\rm d}^8 z  
\lim_{z' \to  z} \, 
\hat{\bG}^{(e)}(z,z')\Big( \, \sum_{\iu ,\ju} (E_{0 \iu} \, E_{\iu 
0})^{0 \ju}{}_{0 \ju} \, \check{\bG}^{(e)}(z,z') + 
\sum_{\iu ,\ju} (E_{0 \iu} \, E_{\iu 
0})^{\ju 0}{}_{\ju 0} \, \check{\bG}^{(-e)}(z,z') \nonumber \\
& & \quad + \, \sum_I \,\sum_{\iu } (E_{0 \iu} \, E_{\iu 
0})^{I}{}_{I} \, \check{\bG}^{(0)}(z,z') + 
\sum_{\iu ,\ku ,\lu} (E_{0 \iu} \, E_{\iu 
0})^{\ku \lu}{}_{\ku \lu} \, \check{\bG}^{0)}(z,z')\, \Big)~.
\eea
This can be rearranged in the form
\bea
\G_\infty^{(1)} &=& \int {\rm d}^8 z  
\lim_{z' \to  z} \, 
\hat{\bG}^{(e)}(z,z') \, \Big( \,\sum_{\iu} {\rm tr}_{\rm Ad}(E_{0 \iu} \, E_{\iu 
0}) \, \check{\bG}^{0)}(z,z') \nonumber \\
& & \quad + \, \sum_{\iu ,\ju} (E_{0 \iu} \, E_{\iu 
0})^{0 \ju}{}_{0 \ju} \, (\check{\bG}^{(e)}(z,z') - \check{\bG}^{(0)}(z,z'))
\nonumber \\
& & \quad + \, \sum_{\iu ,\ju} \, (E_{0 \iu} \, E_{\iu  0})^{\ju 0}{}_{\ju 0} \, 
(\check{\bG}^{(- e)}(z,z') - \check{\bG}^{(0)}(z,z')) \, \Big)~.
\eea
Using ${\rm tr}_{\rm Ad}(E_{i j} \, E_{k 
l})  = 2 N \delta_{i l}\, \delta_{j k} $,
the relation $\lim_{z' \to  z} 
\check{\bG}^{(e)}(z,z') 
=  \lim_{z' \to  z} 
\check{\bG}^{(-e)}(z,z'),$ and the identities (\ref{id5}) 
and (\ref{id8}), one obtains
\be
\G_\infty^{(1)} = N (N-1) \int {\rm d}^8 z  
\lim_{z' \to  z} \, 
\hat{\bG}^{(e)}(z,z') \, \Big( \,\check{\bG}^{(e)}(z,z') + 
\check{\bG}^{(0)}(z,z') \,  \Big)~.
\ee

In a similar manner, $\G_\infty^{(2)},$ $\G_\infty^{(3)}$ and 
$\G_\infty^{(4)}$ can be computed:
\bea
\G_\infty^{(2)} &=& N (N-1) \int {\rm d}^8 z  
\lim_{z' \to  z} \, 
\hat{\bG}^{(-e)}(z,z') \, \Big( \,\check{\bG}^{(-e)}(z,z') + 
\check{\bG}^{(0)}(z,z') \,  \Big)~, \\
\G_\infty^{(3)} &=& 2 (N-1) \int {\rm d}^8 z  
\lim_{z' \to  z} \, 
\hat{\bG}^{(0)}(z,z') \non \\
&& \qquad \qquad \times 
\Big( \,\check{\bG}^{(e)}(z,z') + 
\check{\bG}^{(-e)}(z,z') + 
(N-2) \, \check{\bG}^{(0)}(z,z')\,  \Big) ~,\\
\G_\infty^{(4)} &=& (N-1) (N-2) \int {\rm d}^8 z  
\lim_{z' \to  z} \, 
\hat{\bG}^{(0)}(z,z') \non \\
&& \qquad \qquad 
\times  \Big( \,\check{\bG}^{(e)}(z,z') + 
\check{\bG}^{(-e)}(z,z') 
+ \,2 (N-1) \, \check{\bG}^{(0)}(z,z')\,  \Big)~.
\eea
Combining these results with that for $\G_\infty^{(1)}$ yields 
the required expression (\ref{eight-structure-0})
for  $\G_\infty$.

\end{appendix}

\end{document}